\documentclass{aa}
\usepackage{natbib}               
\usepackage{graphicx}             
\usepackage{amssymb}              
\usepackage[usenames,dvipsnames]{color} 
\usepackage{txfonts}

\newcommand{\BE}{\begin{equation}}
\newcommand{\EE}{\end{equation}}
\newcommand{\BA}{\begin{eqnarray}}
\newcommand{\EA}{\end{eqnarray}}
 \newcommand{\fig}[1]{Fig.~\ref{fig_#1}}

 \newcommand{\figss}[2]{Figs.~\ref{fig_#1} - \ref{fig_#2}}
 \newcommand{\sect}[1]{Sect.~\ref{sect_#1}}
 \newcommand{\sects}[2]{Sects.~\ref{sect_#1} and~\ref{sect_#2}}
 \newcommand{\eq}[1]{Eq.~(\ref{eq_#1})}
 \newcommand{\eqs}[2]{Eqs.~(\ref{eq_#1}) and (\ref{eq_#2})}

\newcommand{\eg}{e.g.}

\newcommand{\ie}{i.e.}
\newcommand{\insitu}{in situ}

\newcommand{\commonCap}{The drawing convention and added parameters are the same than in \fig{Vmax_shock}.}

\newcommand{\degree}{\ensuremath{^\circ}}
\newcommand{\grad}{ {\bf \nabla } }

\newcommand{\rmd}{{\rm d }}

\newcommand{\uvec}[1]{\hat{\bf #1}}

\newcommand{\Baxis}{B_{\rm axis}}
\newcommand{\Bicme}{<$B_{\rm icme}$>}
\newcommand{\bSa}{b/a}
\newcommand{\cp}{c_{\rm p}} 
\newcommand{\cs}{c_{\rm s}} 
\newcommand{\diff}{\rm{diff}}  
\newcommand{\dtsh}{\Delta t_{\rm sheath}}

\newcommand{\kms}{km.s$^{-1}$}
\newcommand{\lA}{\lambda}
\newcommand{\lAmc}{\lambda_{\rm MC}}
\newcommand{\lAsh}{\lambda_{\rm shock}}

\newcommand{\nb}{n_{\rm b}}

\newcommand{\phimax}{\varphi_{\rm max}}
\newcommand{\phimaxmc}{\varphi_{\rm max, MC}}
\newcommand{\phimaxsh}{\varphi_{\rm max, shock}}
\newcommand{\pA}{\varphi}
\newcommand{\pmc}{p_{\rm MC}}
\newcommand{\Pl}{\mathcal{P}(\lambda)}
\newcommand{\pobs}{\mathcal{P}_{\rm obs}}
\newcommand{\pobsl}{\mathcal{P}_{\rm obs}(\lA)}

\newcommand{\Rmean}{R_{\rm mean}}
\newcommand{\Sicme}{S_{\rm icme}}
\newcommand{\Ssh}{S_{\rm sh}}

\newcommand{\ur}{\hat{\bf u}_{\rho}}

\newcommand{\ux}{\hat{\bf u}_{\rm x}}

\newcommand{\Vmax}{V_{\rm max}}
\newcommand{\Vmean}{V_{\rm mean}}
\newcommand{\Vwind}{V_{\rm wind}}

\begin{document}

\title{Quantitative model for the generic 3D shape of ICMEs at 1~AU}

\titlerunning{Quantitative model for the generic 3D shape of ICMEs}
\authorrunning{D\'emoulin et al.}

\author{P. D\'emoulin\inst{1}, M. Janvier\inst{2}, J.J. Mas\'{i}as-Meza\inst{3} \and S. Dasso\inst{4,5}}
   \offprints{P. D\'emoulin}
\institute{
$^{1}$ Observatoire de Paris, LESIA, UMR 8109 (CNRS), F-92195 Meudon, France, \email{Pascal.Demoulin@obspm.fr}\\
$^{2}$ Institut d'Astrophysique Spatiale, UMR8617, Univ. Paris-Sud-CNRS, Universit\'e Paris-Saclay, B\^atiment 121, 91405 Orsay Cedex, France \email{mjanvier@ias.u-psud.fr}\\
$^{3}$ Departamento de F\'\i sica and IFIBA, Facultad de Ciencias Exactas y Naturales, Universidad de Buenos Aires, 1428 Buenos Aires, Argentina \\
$^{4}$ Instituto de Astronom\'\i a y F\'\i sica del Espacio, UBA-CONICET, CC. 67, Suc. 28, 1428 Buenos Aires, Argentina, \email{sdasso@iafe.uba.ar} \\
$^{5}$ Departamento de Ciencias de la Atm\'osfera y los Oc\'eanos and Departamento de F\'\i sica, Facultad de Ciencias Exactas y Naturales, Universidad de Buenos Aires, 1428 Buenos Aires, Argentina, \email{dasso@df.uba.ar}\\
}
   \date{Received ***; accepted ***}

   \abstract  
   {Interplanetary imagers provide 2D projected views of the densest plasma parts of interplanetary coronal mass ejections (ICMEs) while \insitu\ measurements provide magnetic field and plasma parameter measurements along the spacecraft trajectory, so along a 1D cut.  As such, the data only give a partial view of their 3D structures.
   }
   {By studying a large number of ICMEs, crossed at different distances from their apex, we develop statistical methods to obtain a quantitative generic 3D shape of ICMEs. 
   }
   {In a first approach we theoretically obtain the expected statistical distribution of the shock-normal orientation from assuming simple models of 3D shock shapes, including distorted profiles, and compare their compatibility with observed distributions.  In a second approach we use the shock normal and the flux rope axis orientations, as well as the impact parameter, to provide statistical information across the spacecraft trajectory. 
   }
   {The study of different 3D shock models shows that the observations are compatible with a shock symmetric around the Sun-apex line as well as with an asymmetry up to an aspect ratio around 3. Moreover, flat or dipped shock surfaces near their apex can only be rare cases.
Next, the sheath thickness and the ICME velocity have no global trend along the ICME front.
Finally, regrouping all these new results and the ones of our previous articles, we provide a quantitative ICME generic 3D shape, including the global shape of the shock, the sheath and the flux rope.
   }
   {The obtained quantitative generic ICME shape will have implications for several aims. For example, it constrains the output of typical ICME numerical simulations.  It is also a base to develop deeper studies of the transport of high energy solar and cosmic particles during an ICME propagation, as well as for modeling and forecasting space weather conditions near Earth.
      }


   \maketitle

\section{Introduction} 
\label{sect_Introduction}

  The coronal magnetic field recurrently accumulates magnetic field and free magnetic energy as a consequence of 
  photospheric plasma motions and flux emergence.  
  Through the process of photospheric cancellation of magnetic polarities, a sheared arcade is typically transformed into a twisted flux tube or flux rope (FR). At some point during its evolution, the FR can become unstable, erupting upward, with fast reconnection occuring underneath its structure, leading to a flare \citep[e.g., see the reviews of][]{Forbes06,Janvier15a,Schmieder15}.  If the downward magnetic tension of the overlying arcade is not strong enough the flux rope is ejected, creating a coronal mass ejection (CME).
   
\begin{figure}  
\centering
\includegraphics[width=0.5\textwidth,clip]{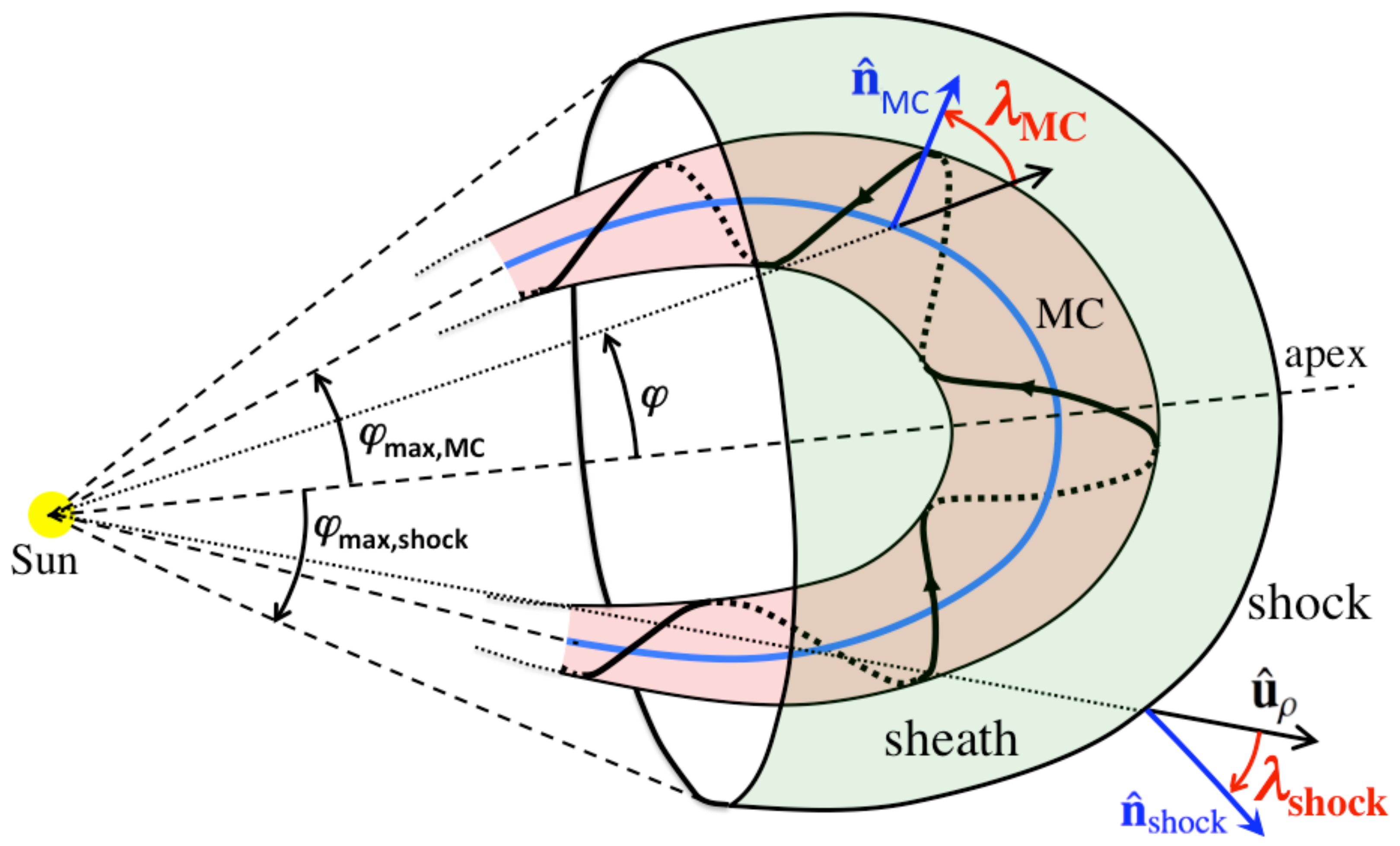}
\caption{Schema of a MC, its sheath and its front shock.  The MC axis is drawn in blue and its maximum angular extension as viewed from the Sun is {2}$\phimaxmc$ while the shock extends up to {2}$\phimaxsh$. $\lA$ is defined by the angle between the radial direction from the Sun and the axis or the shock normal. 
}
\label{fig_schema3D}
\end{figure}

Observations of CMEs by coronagraphs are extended in the interplanetary space with the heliospheric imagers of the STEREO spacecraft \citep[e.g., see the review of][]{Rouillard11b}.    
The interplanetary CMEs (ICMEs) are also detected \insitu\ as an altered solar wind structure \citep[e.g., see the review of][]{Zurbuchen06}. 
Many differences have been found between ICMEs and ordinary solar wind properties: 
a low proton temperature \citep[less than half of what is expected for a typical solar wind with similar velocity,][]{Richardson95,Elliott05,Demoulin09}, an enhanced and coherent magnetic field \citep[\eg][]{Burlaga90}, enhanced ion charge states \citep{Lepri01,Lepri04}, an increase of bi-directional fluxes of suprathermal particles \citep[\eg][]{Marsden87,Gosling87}, etc.  From the plasma parameters and the magnetic field reported in previous and recent ICME studies (and including the present one), a global picture for the structure of ICMEs has emerged. 
This is summarized in \fig{schema3D} with three key ingredients: the magnetic cloud (MC), the sheath and the front shock.

    A MC is identified by an enhanced magnetic field intensity, a large scale and coherent magnetic field rotation, and proton temperatures lower than in the typical solar wind with the same radial speed \citep[\eg][]{Burlaga81,Dasso05}.  These observations are typically interpreted as the crossing of a FR, which corresponds to the unstable magnetic configuration ejected from the Sun,  few days before its detection in the interplanetary medium \citep[\eg\ see the review of][]{Demoulin08}.  MCs are typically present in a fraction (about one third) of ICMEs \citep[e.g.][]{Richardson10}. This is a lower limit, since the other fraction of cloud-like or no MC detected events could be due to a spacecraft crossing too far away from the nose or the centre of the MC (making both the detection and the data fitting difficult). 
Also, the propagation process of MCs in the solar wind can also play a role in making the detection of MCs difficult. For example, the FR erosion upon reconnection with the solar-wind magnetic field creates an asymmetric FR with part of the original FR connected to the solar wind \citep{Dasso06, Ruffenach15}.

Shocks are formed in front of ICMEs/MCs when they travel faster than the encountered solar wind, with a relative speed larger than the faster MHD mode speed \citep[\eg][]{Bothmer98}.   
At 1~AU shocks are frequent in front of ICME sheaths.  For example, shocks were present in front of 64\% of the MCs studied by \citet{Feng10} during the period 1995 to 2007.  It is commonly thought that the shock is driven by the FR present behind \citep[\eg][]{Lario01,Oh02,Marubashi07}.
At 1~AU the shocks have a moderate strength with Mach number and density compression ratio both with a mean just above 2 \citep{Oh07,Wang10}.

ICMEs are typically preceded by a sheath of compressed plasma and magnetic field collected on the way from the Sun \citep[\eg][]{Gosling87b,Demoulin10,Richardson11}.  A shock is typically present at the sheath front border where both the plasma and the magnetic field are suddenly compressed.  The sheath has typically a magnetic field strength comparable, or even larger, than the following MC \citep[\eg\ ][]{Zurbuchen06,Masias-Meza16}.  In the sheath, the magnetic field rapidly changes its orientation (this is due to different solar wind structures pressed together) and the proton temperature is typically much higher than in the solar wind and even more than in MCs (due to a strong compression and an enhanced heating).

  Knowing the FR, sheath and shock shapes, and more generally the whole ICME structure, is important for several reasons.  For example, they set global constraints on the results of numerical simulations.
Also, they are important information for possible impact prediction with the Earth's magnetosphere, their arrival time as well as their journey in the interplanetary medium \citep[\eg][]{Mostl13,Mostl15}. The ICME structure also affects the transport of energetic particles in the heliosphere over a large range of energies \citep[\eg][]{Cane00}.
First, the global shape of MCs, as well as their amount of twist, determine the effective length for the travel of charged relativistic particles, from the Sun to the Earth, producing delays in the observation of ground level enhancements, with respect to the expected time for a Parkerian solar wind \citep[\eg][]{Kahler11,Masson12,Hu15}. 
Second, at higher energies (galactic cosmic rays), the passage of an ICME frequently modifies the stationary flux of particles arriving to the Earth surface \citep[\eg\ producing a Forbush decrease,][]{Cane00}, which can be observed from ground-level instruments, as neutron monitors \citep[\eg][]{Simpson54,simpson00} or water Cherenkov detectors \cite[\eg][]{auger11,Dasso12,lago16}.
Finally, at lower energies, ICMEs can produce changes in fluxes of suprathermal particles that can be observed \insitu\ in the heliosphere by spacecraft \cite[\eg][]{Mulligan09}. 
All in all, a better knowledge of the ICME generic structure helps the progress on ICME modelling and forecasting for space weather.

In a series of papers, the ICME generic structure was characterized from \insitu\ data of large sample of events.   
\citet{Janvier13} defined this new statistical method and deduced the mean shape of MC axis from the local FR axis orientation of a MC set based on the results of \citet{Lepping10}. They introduced the location angle $\lA$ defined as the angle between the radial direction from the Sun and the local normal to the MC axis (\fig{schema3D}).
$\lA$ defines the location of the spacecraft crossing along the flux-rope axis when its shape is known. The generic MC axis was deduced from the observed probability function, $\Pl$. 
They found an axis shape close to an ellipse shape with an aspect ratio of $1.2$ (lower extension in the radial direction).  
   The above statistical method was extended to shocks present in front of ICMEs by \citet{Janvier14b} with $\lA$ defined as the angle between the radial direction from the Sun and the local normal to the shock (\fig{schema3D}). 
   They showed that \insitu\ data could be compatible with an axisymmetric shock shape around the direction Sun to shock apex.  They also showed that these statistical results are in agreement with imager data of a well observed ICME \citep{Mostl09c}.
  Next, while the relative directions of MC axis and shock normal have a large dispersion when analysed case by case \citep{Feng10}, 
  the above statistical method applied to the same sample of cases showed comparable results for the MC axis and shock shapes \citep{Janvier15b}.  
  More generally, they show that MC axis and shock normals lead to comparable ellipsoidal shapes with all the \insitu\ data sets analysed. 
  Finally, the ellipsoidal model has the closest probability, $\Pl$, from the observed distributions for the three models tested.  
       
In this paper we continue the effort to determine the ICME generic structure by first exploring the effect of various shock surface shapes, and in particular some distortion effects, on the distribution of $\lA$ (\sect{Model}).  
The comparison of this prediction to the observed distribution from shock normals sets some constraints on the generic shock structure.  
We next explore the possible observational constraints which can be set on ICME sheaths from \insitu\ data (\sect{Sheath}). 
Then, we combine the previous results with those obtained earlier on to derive a global generic and quantitative structure of ICMEs (\sect{ICME}). 
Finally, in \sect{Conclusion}, we first summarize the main results, conclude and outline potential applications.     
   
\section{Testing different shock shapes with in situ data} 
\label{sect_Model}

  In this section we explore the implications of various 3D shock shapes and distortions effects on the probability distributions of the location angle $\lA$ (defined in \fig{schema3D}).  The main aim is to explore whether various shock shapes are compatible with the observed distribution of $\lA$,  and whether there could be generic shock shapes in front of ICMEs.    
   
\subsection{3D shock shape} 
\label{sect_Mod_3D}

In order to simplify the description of the 3D shock shape, the model is described in Cartesian coordinates $(x,y,z)$ with the $x$ coordinate being the radial direction (direction of propagation) away from the Sun and $y,z$ two orthogonal directions.  
The shock surface in general can be described by an equation of the type $f(x,y,z)=0$.  
The shock normal $\uvec{n}=(n_x,n_y,n_z)$ is along $\grad f$.  The location angle, $\lA$, is related to $\uvec{n}$ with    
  \BE \label{eq_tanLambda}
  \tan \lA =  \sqrt{n_y^2+n_z^2} \, \bigg/ \, n_x \, .
  \EE
Here $\lA$ is defined with respect to the fixed direction $\ux$, while in previous studies it was defined with respect to $\ur$ (\fig{schema3D}). This is an approximation better suited for narrower shocks (see below).    

Next, we compute the probability of having $\lA \pm \rmd \lA/2$ during a spacecraft crossing the shock of a large number of CMEs.  We suppose a uniform probability of the spacecraft location in the $y,z$ plane.  
This uniform probability is supposed since a spacecraft typically observes a large number of ICMEs launched from the Sun from a broad range of latitudes and longitudes (\sect{Mod_obs}). Similarly, the ACE spacecraft crossed over several years ICME shocks with no privileged location.  Numerically, we set a uniform grid in $y,z$, and compute $\lA$ from \eq{tanLambda}. 
The probability density function (PDF) of $\lA$ is then computed by normalizing the histogram of $\lA$ by the sum of bin counts.

\begin{figure}  
\centering
\includegraphics[width=0.3\textwidth,clip]{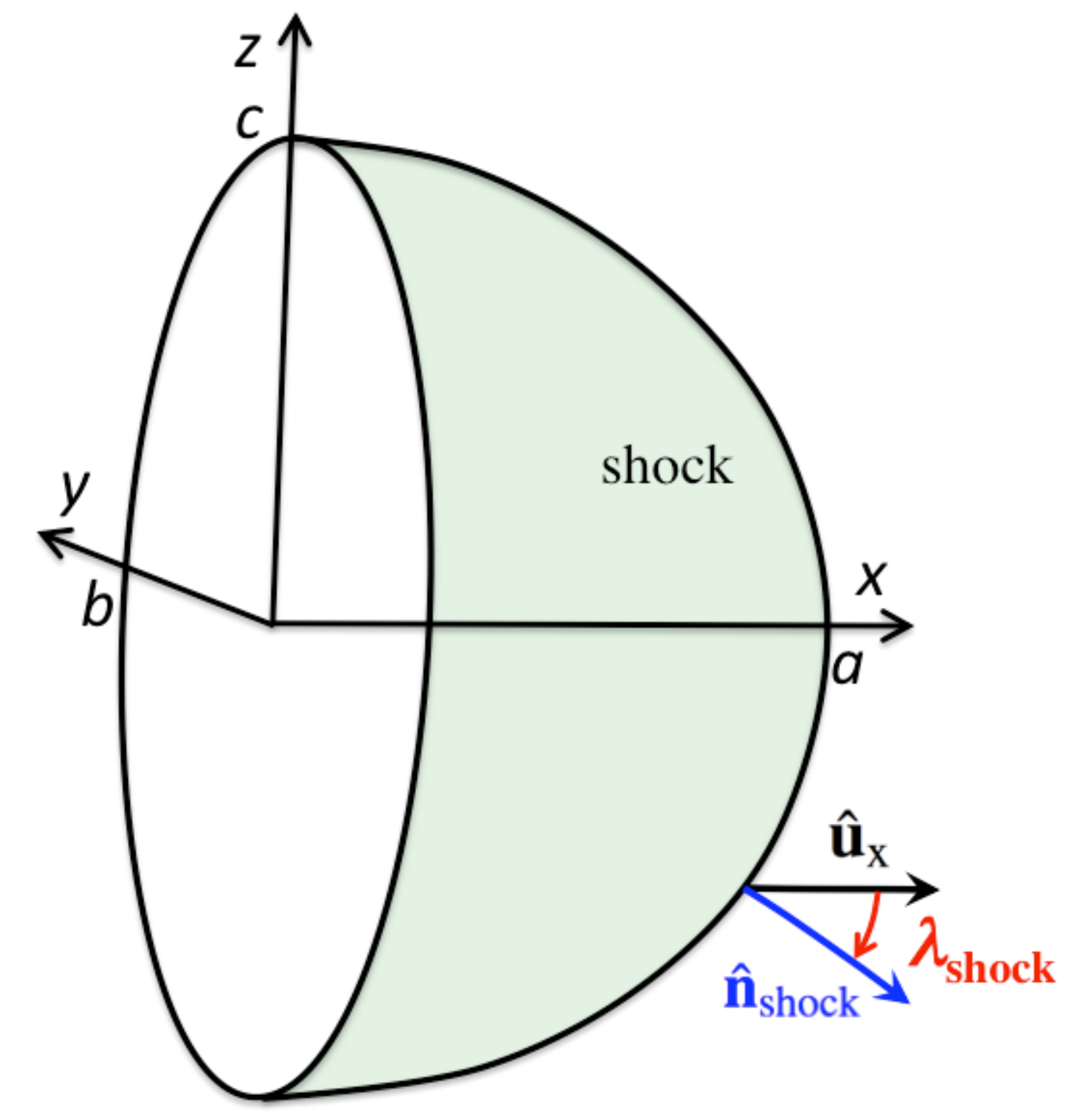}
\caption{Schema of the front shock in Cartesian coordinates.  $x$ is the propagation direction of the following ICME (not represented).  $\lA$ is defined by the angle between the $x$ direction and the shock normal. 
}
\label{fig_schemaCart}
\end{figure}

\subsection{3D ellipsoidal model} 
\label{sect_Mod_Ellipsoidal}

 The shock front is described by the $x>0$ half part of an ellipsoidal surface, shown in \fig{schemaCart}:
  \BE \label{eq_f}
  f(x,y,z) = (x/a)^2 + (y/b)^2 + (z/c)^2 -1 = 0 \, .
  \EE
$\tan \lA$, derived from \eq{tanLambda}, is a function of $y,z$ and of the shock shape as:
  \BE \label{eq_tanLambda_ellip}
  \tan \lA = \sqrt{\frac{(a~y/b^2)^2+(a~z/c^2)^2}
                  {1 - (y/b)^2 - (z/c)^2} } \, .
  \EE
In general, the analytical derivation of the probability $\Pl$ is not simple, then we computed it numerically.   
Two limits are still easily derived analytically, as follows.

\begin{figure}  
\centering
\includegraphics[width=0.5\textwidth,clip]{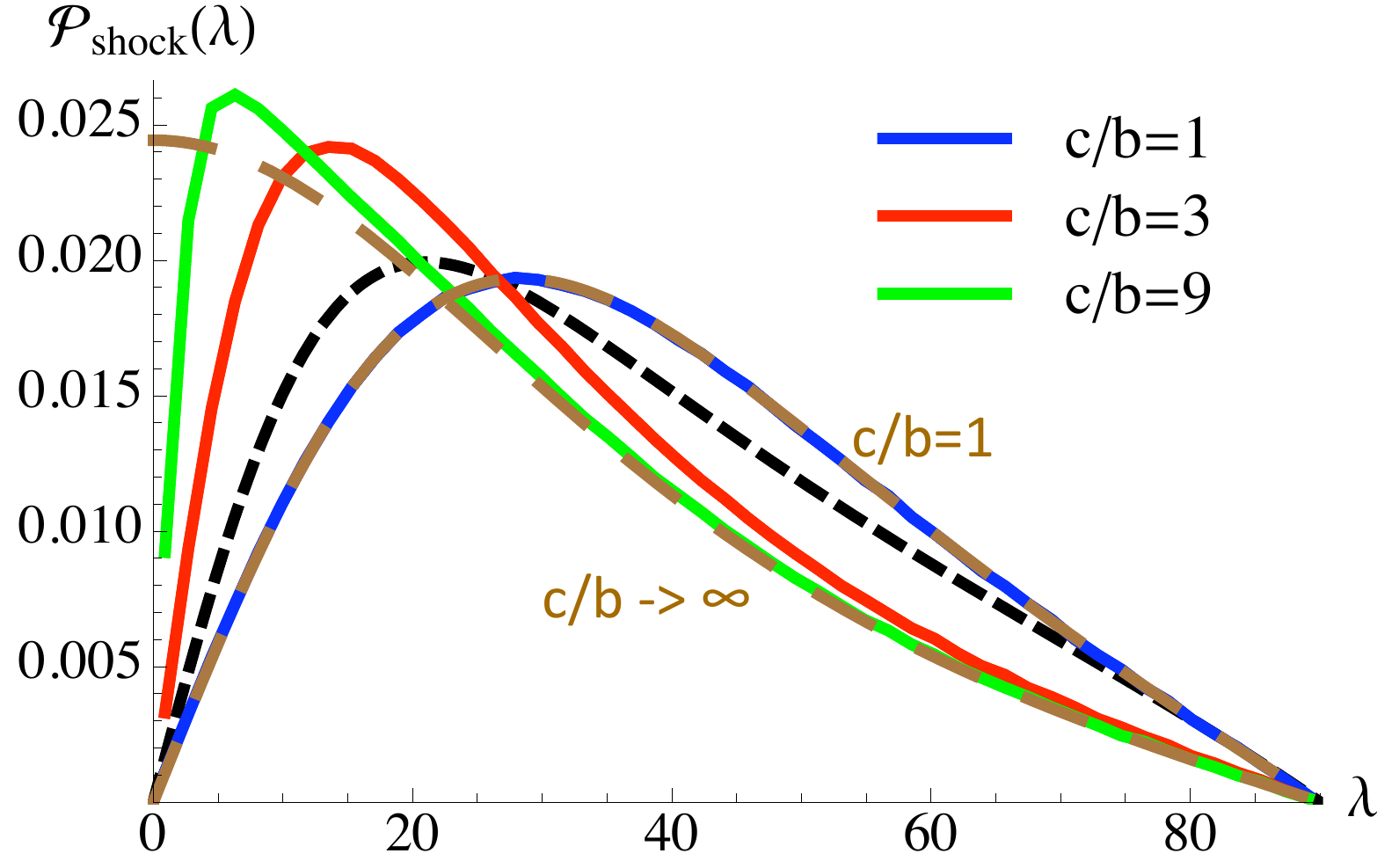}
\caption{Probability of $\lA$, $\Pl$, for the 3D elliptical model, \eq{f}, for different aspect ratio, $c/b$, defined in the plane orthogonal to the shock propagation direction (see \fig{schemaCart}).  The two dashed brown curves are the two analytical limits of \eqs{Pl_cart_cyl}{Pl_cart_y}.  The black dashed curve is the axisymmetric model of \citet{Janvier15b} with a total angular width $2\phimax = 30\degree$. The difference between the black and blue-brown dashed curves is that $\lA$ is computed from the local radial direction from the Sun for the first one and from the $x$-direction for the second (see \sect{Mod_Ellipsoidal}).
For all curves $b/a=1.4$. 
}
\label{fig_ellipse_cart}
\end{figure}

For the case with axi-symmetry, \ie , $b=c$, $\lA$ is simply a function of $r$ with
$r=\sqrt{y^2+z^2}$.  The probability $\Pl ~\rmd~\lA$ is proportional to the surface $2 \pi r ~\rmd~r$.  The fraction $\rmd \lA/\rmd r$ is computed by deriving \eq{tanLambda_ellip}. Finally, $\Pl$ writes:
  \BE \label{eq_Pl_cart_cyl}
  \Pl = \frac{2 ~(b/a)^2 \sin \lA \cos \lA}
                  {\Big( (b/a)^2 \sin^2 \lA + \cos^2 \lA \Big)^2}  \, .
  \EE
A second analytical case is found in the limit $c \rightarrow \infty$, so when the cross section of the shock in the $xy$-plane extends indefinitely in the $z$-direction (i.e., a semi-elliptic cylinder with its axis along the $z$-direction).  In this limit, we derive:
  \BE \label{eq_Pl_cart_y}
  \Pl = \frac{b/a  \cos \lA}
                  {\Big( (b/a)^2 \sin^2 \lA + \cos^2 \lA \Big)^{3/2}}  \, .
  \EE
These cases are used to test the numerical computations of $\Pl$ (brown dashed curves in \fig{ellipse_cart}).

Next, we compare $\Pl$ of the axisymmetric case, $b=c$ to the axisymmetric model of \citet{Janvier15b}.  This last model was derived in spherical geometry, \fig{schema3D}, while the present model is derived in Cartesian geometry, \fig{schemaCart}.  The difference is that the reference direction $\ur$ to compute $\lA$ is changing of direction along the shock (\fig{schema3D}), while $\ux$ has a fixed direction. Then, the difference to compute $\lA$ between the models in spherical and Cartesian geometries is the angle between $\ur$ and $\ux$.  
When the shock is crossed at the nose, the radial direction (spacecraft crossing) is the same as the propagation direction, so that $\ur$= $\ux$. If the spacecraft is crossed further away from the nose, then the radial direction becomes quite different from the propagation direction, especially if the maximum angular extension of the shock $\phimax$, defined in \fig{schema3D}, is large.  For small values of $\phimax$, {$\ur\approx \ux$, and as such the 3D ellipsoidal model is suited for narrow ICMEs.
\fig{ellipse_cart} shows that $\Pl$ distributions are still comparable for a spherical model  with a total angular width of $30 \degree$ (black dashed line) and the 3D ellipsoidal model (blue line).  This last numerical result is identical to \eq{Pl_cart_cyl} (brown dashed line). 

 Finally, we analyze the 3D ellipsoidal model to explore the effect on $\Pl$ of modifying the shock shape to non-axisymmetric configurations.
Increasing the ratio $c/b$ above 1 implies a flatter surface.  For a given $b/a$ \citep[taken as 1.4, a value close to the ones derived from previous studies, see][]{Janvier15b,Mostl15}, this implies a shift of the peak in the $\Pl$ distribution to lower $\lA$ values as $c/b$ increases (\fig{ellipse_cart}).  In the limit of large $c/b$ ratio, $\uvec{n}$ is mostly located in the $x,y$ plane, and the distribution of $\lA$ is defined by the curvature of a surface nearly invariant in the $z$ direction. Because of this invariance, the distributions of $\lA$ become similar to those found for a curve related to the same shape lying in the x-y plane. 
Indeed, this distribution of $\lA$ is close to the one found for the MC axis \citep{Janvier13}, while being significantly different from what is found for the shocks \citep{Janvier14b}.  

\begin{figure*}  
\centering
\sidecaption
\includegraphics[width=0.7\textwidth,clip]{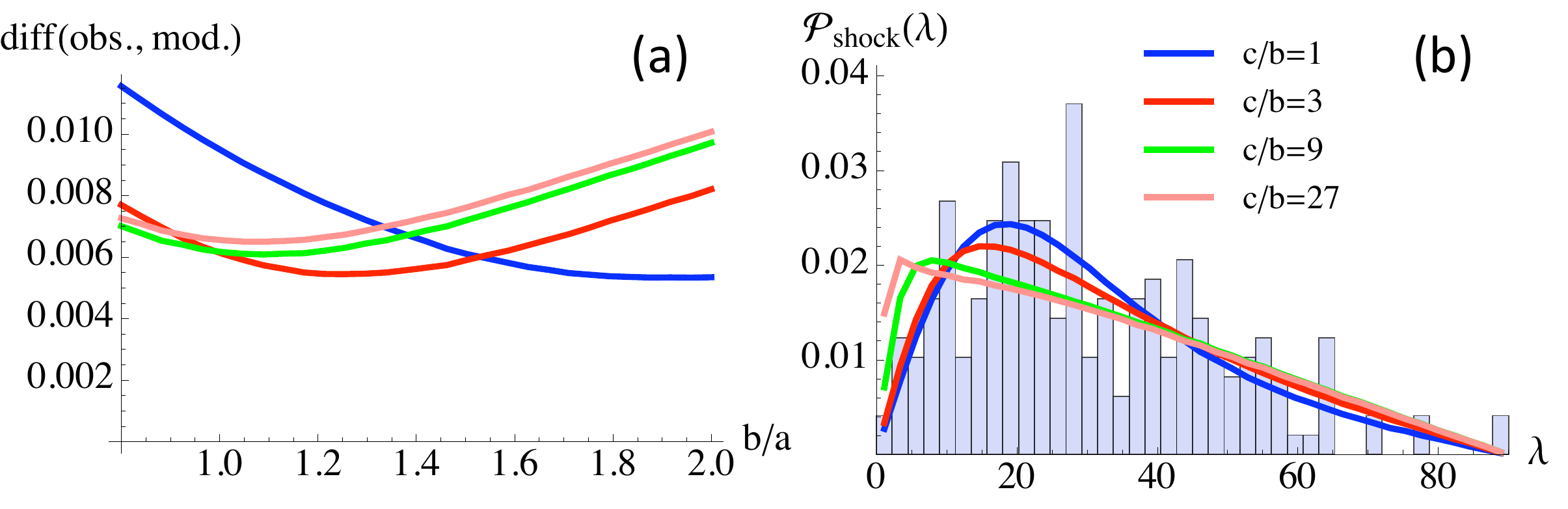}
\caption{Minimisation of the difference between shock observations \citep[blue histogram,][]{Wang10} and the 3D ellipsoidal model (\sect{Mod_3D}).(a) The difference function is defined by \eq{diff} and it is plotted in function of the aspect ratio $\bSa$.  (b) The theoretical probability, for the shock shape described by \eq{f} and least squared fitted to the observed distribution, is shown for four values of the aspect ratio $c/b$ ($=1$ for the axisymmetric case).  Here, $b/a$ is defined by the minimum of $\diff {\rm (obs., mod.)}$ of panel (a) with the curve of the same color. 
}
\label{fig_ellipse_obs}
\end{figure*}

\subsection{Comparison to observations} 
\label{sect_Mod_obs}

Below, we compare the 3D ellipsoidal model to the \insitu\ data of shocks studied by \citet{Wang10}.  This study extends from February 1998 to August 2008 and the authors have analysed 216 shocks. They derived the shock parameters, in particular the shock normal, by fitting the MHD Rankine-Hugoniot relations to the data in the vicinity of the shocks.  117 shocks are in front of ICMEs as identified by \citet{Richardson10} and the 99 remaining shocks have no detected ICMEs behind.  

A small fraction of these last shocks is due to a fast solar wind stream overtaking a slower solar wind (stream interaction regions or SIRs). These shocks may represent as much as 20-30\% of shocks measured at 1~AU during the time period 1995 to 2004 \citep{Jian06b}.
Around solar maximum and at 1~AU such shocks are about 10 times less numerous than ICME shocks \citep{Lai12}. Indeed \citet{Janvier14b} found comparable distributions $\Pl$ for shocks with and without an associated detected ICME, and as such the deduced mean shock shapes are very close.  Of course this similarity of $\Pl$ could also indicate that the physical processes involved in ICME and SIR shocks are similar because for both cases the driven structure (ICME or fast wind) is mainly travelling radially from the Sun and overtaking a slower plasma. However, we follow below the interpretation of \citet{Janvier14b} who concluded that most shocks observed at 1~AU are in fact associated with an ICME, although the ICME is not always detected as the shock has a broader spatial extension than the ICME. Then, we use below all the shock data set to derive the observed $\Pl$, called $\pobsl$. The conclusions are similar using $\Pl$  only from shocks associated to ICMEs.

The analytical model described in \sect{Mod_Ellipsoidal} provides a continuous function of $\lA$ (\eg\ \eqs{Pl_cart_cyl}{Pl_cart_y}). To compare this function with the observed probabilities, we first need to bin this analytical function, similarly as what is done for the observations (the probability function of this binned model is noted $\mathcal{P}_{\rm bm} (\lA )$ in the following).  We also express both probabilities with the same unit, per unit of degree.  The difference between the observed probability and the analytical one is quantified by computing the least square difference with    
  \BE \label{eq_diff}
  \diff {\rm (obs., mod.)} = \sqrt{ \frac{1}{\nb} \sum_{i=1}^{\nb} 
                 \Big(\pobs (\lA_i)-\mathcal{P}_{\rm bm}(\lA_i) \Big)^2                 } \, ,
  \EE
where $\nb$ is the number of bins.

We explore below how the ratio $c/b$ affects the fit of the modeled $\Pl$ to the observations by finding the ratio $b/a$ which minimizes $\diff {\rm (obs., mod.)}$ for a fixed $c/b$ value. We find that the maximum of $\Pl$ shifts to lower $\lA$ values for larger $c/b$, \fig{ellipse_obs}b, as in \fig{ellipse_cart}.
However, since a least square fit is made to a fixed observed distribution (blue histogram in \fig{ellipse_obs}), the effect of increasing $c/b$ is softened by a reverse change of $b/a$ (\fig{ellipse_obs}a). Decreasing values of $b/a$ means that the shock shape is more bended in the $xy$-plane, then $\Pl$ decreases for smaller $\lA$ values and increases at larger $\lA$ values, then moderating the $\Pl$ changes outlined in \fig{ellipse_cart} with an increasing $c/b$. 
Still, the values of $\Pl$ are far away from that of $\pobsl$ with increasing $c/b$ values. For $c/b=3,9,27$ the minimum value of $\diff {\rm (obs., mod.)}$ grows by 2, 14, 22 \% of its minimum value for $c/b=1$.  This growth is much larger in cases for which only the $\lA$ values close to the origin are used to compute $\diff {\rm (obs., mod.)}$ as illustrated in \fig{ellipse_obs}b with the larger distance between the curves at small $\lA$ values compared to larger ones.
Then, shock configurations too far from the axisymmetric case are incompatible with observations. 
However, the upper limit set on $c/b$ is large as the case $c/b=3$ fits the observed distribution nearly as well as the case $c/b=1$ (\fig{ellipse_obs}).
This result is compatible with the equivalent ratio deduced from coronagraph observations of CMEs typically imaged at solar distances below half AU:  \citet{Cremades05} and  \citet{Cabello16} deduced $c/b \approx 1.6$ both from single perspective observations performed on different CMEs and from a single CME observed in quadrature. 

   We conclude that the observed distribution $\pobsl$ could still be compatible with front shocks which depart significantly from axisymmetry while coronagraph observations closer to the Sun point to a low asymmetry.

\begin{figure*}  
\centering
\includegraphics[width=\textwidth,clip]{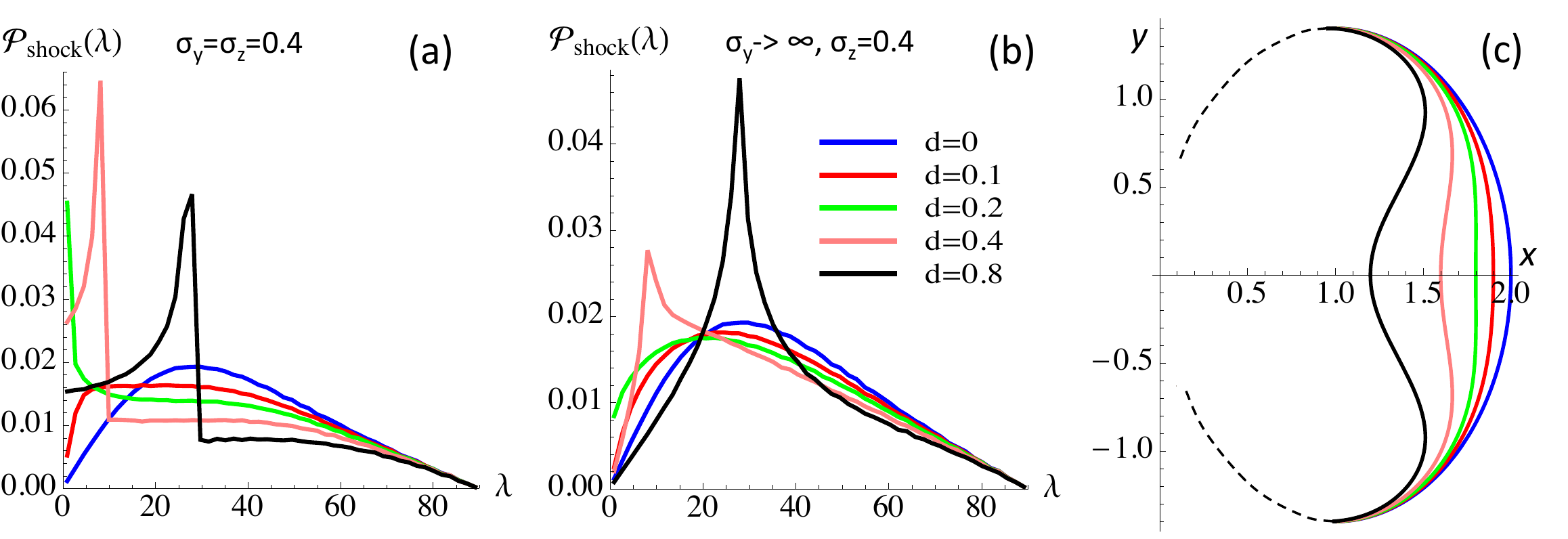}
\caption{(a,b) Theoretical distributions obtained with a 3D ellipsoidal model having a variable dip shape at its front as defined by \eq{x(y,z)_dip} and shown in (c) with a cut in the $x$-$y$ plane. The color curves are for different dip depths (parameter $d$). (a) the front is asymmetric around the $x$ axis (direction of CME propagation) and (b) the dip has no dependence in $y$. For all curves $b/a=1.4$ and $c/b=1$. 
}
\label{fig_ellipse_dip}
\end{figure*}

\subsection{Models with a front dip} 
\label{sect_Mod_dip}
 
Many numerical simulations of ICMEs show a shock front which have, at least globally, an ellipsoidal shape as modelled above \citep[\eg\ ][]{Manchester04,Xiong06,Jacobs07,Taubenschuss10,Zhou14}.  However, when the flux rope travels in a structured solar wind, with some dense and high plasma $\beta$ regions, the front shock shape can be significantly deformed.  In particular, the shock apex takes a concave-outward shape, or dips, for cases when the flux rope propagates in a slow wind edged on both sides by fast winds \citep{Riley03,Manchester04b,Xiong06,Taubenschuss10}. Such a dipped shape requires a very steady and bimodal solar wind which is not present around solar maximum or with a succession of CMEs \citep[\eg\ ][ respectively]{Shen14, Lugaz07}.  
   
We investigate below the effect of such a shock deformation on the distribution $\Pl$. First we rewrite $x$ explicitly in function of $y,z$ from \eq{f}, then we subtract a Gaussian shape:  
  \BE \label{eq_x(y,z)_dip}
   x/a = \sqrt{1- y'^2 + z'^2} - d\, e^{-(y'/\sigma_y)^2-(z'/\sigma_z)^2} \, ,
  \EE
with $y'=y/b$ and $z'=z/c$.
The depth of the dip is characterized by $d$ and its extension in $y,z$ by $\sigma_y,\sigma_z$, respectively.
For example, with a slow and dense solar wind located around the $x$-$y$ plane, surrounded by a faster solar wind at larger $|z|$ values, the dip is expected to have $\sigma_z << \sigma_y$ with a larger $d$ value for a denser slow solar wind.

We show in \fig{ellipse_dip} the results for two extreme cases: the axisymmetric case in panel (a), and the dipped case which only depends on the $z$ coordinate in panel (b).  They outline the range of possible cases obtained by varying the parameters $d,\sigma_y,\sigma_z$.  Peaks in the distribution functions are present where the shock shape has an inflection point (\fig{ellipse_dip}c) because this implies that $\lA$ has nearly the same value in a large range of $y,z$ values, then $\Pl$ is larger.  This effect is more marked in the axisymmetric case, \fig{ellipse_dip}a, since the effect cumulates around circles of constant radius $\sqrt{y^2+z^2}$. On the other hand, for cases with $\sigma_y$ significantly larger than $\sigma_z$, the inflection point is present on a larger range of $\lA$ values, then the peak of $\Pl$ is broader, especially on its right side (\eg\ see \fig{ellipse_dip}b for the limit $\sigma_y \rightarrow \infty$).

For moderate $d$ values, the shock shape is nearly flat around the apex.  This implies significant values of $\Pl$ near the origin (\eg\ the green curves in \fig{ellipse_dip}a,b).  This contrasts with the observation results which show distributions $\pobsl$ with low values near the origin \citep[histogram in \fig{ellipse_obs}b and see other data sets in][]{Janvier15b}.

We next scan the parameter space of  $\sigma_y,\sigma_z$ and $d$. The results can be summarized as follow.
Decreasing $\sigma_y,\sigma_z$ values brings the peaks of $\Pl$ to lower $\lA$ values (not shown). This can be partly compensated by increasing the dip depth ($d$), so that \fig{ellipse_dip} gives a fair overview of the $\Pl$ distributions obtained by scanning the space of parameters $d,\sigma_y,\sigma_z$.  Then, the results of \fig{ellipse_dip} also describe the typical results that we found with a full exploration of the parameter space (within the limits of plausible shapes, \ie\ those found in numerical simulations). 
 
The results above show that flat and concave-outward shapes are not typical of observed ICME shocks. On the one hand, if a flat front was a common feature of their shapes, then the observed distribution $\pobsl$ would not have small values close to the origin as observed. On the other hand, if a dipped front with a characteristic shape was a common feature, then the observed distribution $\pobsl$ would have a marked peak around some $\lA$ values (which would characterize the dip shape).   It is possible that the peaks at $\lA \approx 10\degree$ and $28\degree$ in \fig{ellipse_obs}b could be due to an enhanced contribution of shocks with dips. However, they could also be due to statistical fluctuations due to the low statistics present in each bin: there are around 14 cases per bin in the maximum region of $\pobsl$, so a statistical fluctuation of $\pm 4$, which implies a probability fluctuating in the range $[0.017,0.033]$, a range that is comparable to what is observed.  That said, the interaction of ICMEs with different slow solar winds and different interacting configurations could smooth the observed distribution by creating a variety of dips. Finally, since various dip shapes imply very different $\Pl$, the observed distribution $\pobsl$ is not sufficient to set an estimation of the proportion of dipped cases, nor to estimate the dip shape distribution.    

The above conclusion that the concave-outward shapes are relatively rare is in agreement with the results of the STEREO spacecraft imagers which show only few observed cases \citep{Savani10,Lugaz11}.   This is also an expected result since
the characteristics of the deformation depend on many parameters, such as the geometry of the interaction, in particular the relative orientation of the flux-rope propagation direction and the slow wind layer region.  Then, the symmetric configurations analysed in the numerical simulations \citep{Manchester04b,Xiong06,Taubenschuss10} are not expected to represent typical ICME cases.

\begin{figure*}  
\sidecaption
\centering
\includegraphics[width=0.6\textwidth,clip]{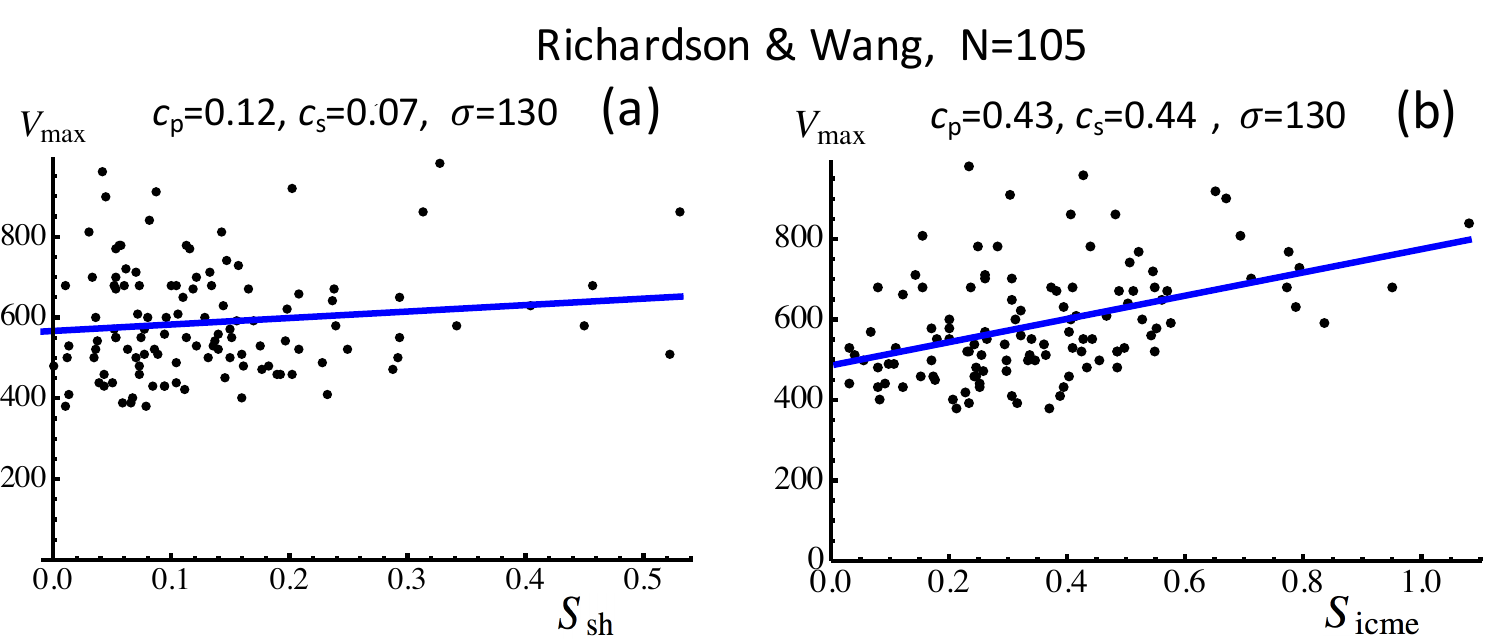}
\caption{Correlations between the maximum velocity, $\Vmax$, within ICMEs and the size of their sheaths, $\Ssh$ (a), and their sizes, $\Sicme$ (b), for the ICMEs studied by \citet{Richardson10} and having a front shock studied by \citet{Wang10}.  The blue straight line is the least square fit to the data points (black points). $\cp$ and $\cs$ are the Pearson and Spearman correlation coefficients, and $\sigma$ is the standard deviation of the ordinate.
}
\label{fig_Vmax_shock}
\end{figure*}

\section{Sheath and ICME properties} 
\label{sect_Sheath}

  In this section we explore the statistical properties of ICMEs in order to get information on the global ICME parameters across the spacecraft trajectory.
We first summarize the data used in \sect{Sheath_association}, then we explore some global properties in \sect{Sheath_global}, before setting the new statistical method used in \sect{Sheath_method} and finally describing the results in \sect{Sheath_along}.  
   
\subsection{Association of ICMEs with Shocks and MCs} 
\label{sect_Sheath_association}

\citet{Richardson10} report on 317 ICMEs observed from 1996 to 2009. In particular, they listed the beginning of the disturbance ($t_{\rm dist}$), associated with the time of the shock, the start ($t_{\rm start}$) and the end ($t_{\rm end}$) times for each detected ICME.
The time difference $\dtsh = t_{\rm start}-t_{\rm dist}$ defines the observed duration of the sheath.  

In order to get complementary information on ICMEs we associate them to shocks or/and MCs when they are observed.  In particular, this provides an estimation of where each ICME was crossed by the spacecraft (see \sect{Sheath_method}).

We associate shocks studied by \citet{Wang10} to the beginning of ICME disturbances of \citet{Richardson10} using a time window of 2 hours following the procedure described in Section 2.3 of \citet{Janvier14b}.
  The Wang's list of shocks, observed by ACE spacecraft, extends from February 1998 to August 2008 and it contains a total of 257 shocks with their main properties (\eg\ shock strength and normal).  This reduces to 117 pairs of associated ICMEs-shocks.

Next, we use an extended list of events (Table~2 at http://wind.nasa.gov/mfi/mag\_cloud\_S1.html) which is based on the fit by a Lundquist's model of the \insitu\ data \citep{Lepping10}. The list contains the parameters obtained for 121 MCs observed by Wind spacecraft from February 1995 to December 2009. Following \citet{Janvier13}, we restrict Lepping's list to 107 MCs (avoiding the cases having the worse model fit to the data).  The association with ICMEs is done by maximizing the overlapping time period between each MC and the ICMEs, with a minimum overlapping interval of 10 hours.  In the few cases where two MCs are associated to one ICME, we selected the MC with the largest common time interval.  We also kept only the MCs crossed by the spacecraft at a distance of the flux-rope axis less than 70\% of its radius (\ie\ the absolute value of the impact parameter $\leq 0.7$) in order to avoid uncertain fitted parameters for too distant encounters. The association and filtering provides 67 pairs of associated ICMEs-MCs.

The above ICMEs associated to a shock can be next associated to MCs with the above procedure.  This restricts the data to 30 triads of associated ICMEs-shocks-MCs.
Since it limits the statistics, we rather present below the results with the pairs ICMEs-shocks and ICMEs-MCs.  The results with the triads are typically comparable while with larger uncertainties due to the limited number of cases.   

\subsection{Sheath global properties} 
\label{sect_Sheath_global}

 We compute the ICME sheath thickness, $\Ssh$, by taking the integral of the radial speed component in the time interval defined by \citet{Richardson10}.  We use the data from OMNI with spacecraft located at 1~AU. 
Some of the ICME sheaths have data gaps, so we can perform the computations on only 268 cases of Richardson and Cane list. 
We find a mean sheath thickness of $0.1$~AU which is comparable to the mean value of $0.082$~AU found by \citet{Richardson10} for the same dataset and $0.11$~AU found by \citet{Mitsakou14} for ICMEs observed during 1996-2008.  

The sheath thickness $\Ssh$ is very case-dependent with values in the range [0, 0.5] AU (\eg\ \fig{Vmax_shock}).  This is an expected result since CMEs near the Sun have a broad range of speeds (from $\approx 30$ to more than $\approx 2000$~\kms, see \citet{Yashiro04,Robbrecht09b,Wang11}), so that an efficient snow-plow effect is highly case-dependent.  Moreover, FRs are expected to reconnect with the overtaken solar wind with a variable efficiency (\eg\ depending on the relative velocities and magnetic field orientations).   Indeed, the percentage of reconnected flux at 1~AU ranges from no significant reconnected flux to at least 60\% in MCs \citep{Dasso06,Dasso07,Mandrini07,Ruffenach12,Ruffenach15}.  This variable amount of reconnection between the FR and the magnetic field accumulated in the sheath is another source of variable sheath thickness.  

At first sight, faster ICMEs are expected to have an average larger sheath as they overcome more solar wind plasma during their travel from the Sun to $D=1$~AU.  For example, with a constant front velocity equal to $\Vmax$, and solar wind velocity,  $\Vwind$, the radial size of the overtaken solar wind is $(1-\Vwind/\Vmax) D$, so a monotonously growing function of $\Vmax$.  However, including the shock properties, as deduced from hydrodynamics physics within a mass budget of the sheath, \citet{Siscoe08} found a sheath thickness, relative to the radius of curvature of the driver, which is a decreasing function of the Mach number. Then, with a comparable overtaken solar wind, the sheath thickness is rather expected to decrease with $\Vmax$.
In fact we find that the sheath thickness is weakly dependent of $\Vmax$ within the dispersion of individual cases (\fig{Vmax_shock}a).  The large dispersion of $\Ssh$, \eg\ due to reconnection, may wash out the expected dependency.  The radius of curvature is also not available from in situ data so we cannot normalize the sheath thickness as in \citet{Siscoe08}. 

For comparison with the sheath, we also compute the ICME size $\Sicme$.  The faster ICMEs are typically larger (\fig{Vmax_shock}b).   A comparable positive correlation, around 0.4 to 0.5, is found between $\Sicme$ and $\Vmax$ in different sub-groups of ICMEs, for example with/without MCs (not shown).  These results are in agreement with the ones of Figure 12 of \citet{Richardson10} who find a positive correlation of the expansion speed with both the radial size and velocity of ICMEs.  Next, what is the origin of this positive correlation between $\Sicme$ and $\Vmax$?  Reconnection at the ICME front removes an equivalent amount of magnetic flux in the front of the FR than in the rear of the sheath, while the corresponding flux stays at the rear of the ICME.
This back region is connected to the solar wind, with a mix of properties between MC and solar wind ones.  In particular, this back region has typically a weaker magnetic field than in the FR \citep{Dasso07,Ruffenach15}, so when magnetic field reconnection is present, the back region is more extended than when it was still part of the FR.  Then, the front reconnection decreases the ICME size at the front while it extends the rear, mitigating the effect of reconnection on the ICME size.  We concluded that the positive correlation $\Vmax (\Sicme)$ is the remaining of an intrinsic correlation, likely not too affected by the amount of front reconnection. 

Finally, we notice the presence of some large $\Ssh$ and $\Sicme$ values in \fig{Vmax_shock}.  The large sheaths may be due to a merging process with the compressed plasma present in a SIR, or they may contain an overtaken ICME which is compressed and hotter. Large ICMEs may be also due to the interaction and the merging of two ICMEs  \citep[\eg ][]{Wang02,Dasso09,Liu14,Lugaz14}.  It is difficult to filter such events without a detailed analysis of each case.
However, such events do not depart from the general tendency (\fig{Vmax_shock}) and they are not numerous so in practice filtering the large events, \eg\ considering only cases with $\Ssh <0.3$ and $\Sicme <0.5$, does not affect significantly the least square fit.  This is also the case in the following analysis (\eg\ as shown in \fig{lA_p_Ssh}).


\begin{figure*}  
\centering
\includegraphics[width=\textwidth,clip]{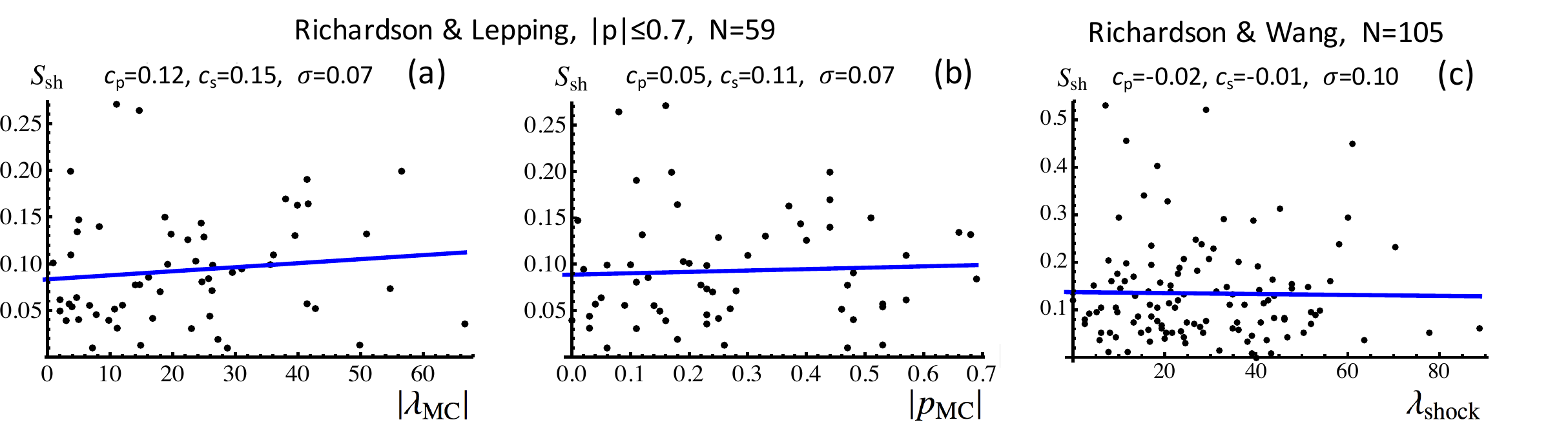}
\caption{Variation of the sheath size, $\Ssh$, along (a) the MC axis ($|\lAmc|$), (b) across the MC axis ($|\pmc|$), and (c) along the shock ($\lAsh$). \commonCap\ Panels (a,b) are for the MCs of Lepping's list and the panel (c) for the shocks studied by \citet{Wang10}. Both sets were associated to the ICMEs studied by \citet{Richardson10}.
The panels show that there is no global dependence of $\Ssh$ value along and across the flux rope as well as along the shock.   }
\label{fig_lA_p_Ssh}
\end{figure*}

\subsection{Statistical method for exploration across spacecraft trajectories} 
\label{sect_Sheath_method}

A single spacecraft explores the crossed ICME only along a line parallel to the main ICME global motion (nearly in the radial direction from the Sun).  Moreover, the location of the crossing within the ICME is typically not known.  Exploring the ICME properties in the ortho-radial directions would need an ensemble of spacecraft distributed over the ICME cross section, which is presently not available, or to have multi-crossing of a single spacecraft, which is impossible as the ICME is too fast and not massive enough to orbit around it.

For most ICME shocks $\lAsh$ is a monotonic function of $\pA$, the angular distance from the shock apex as seen from the Sun (\sect{Mod_dip}). With a shock shape defined, \eg\ with the generic ICME shock shape determined by \citet{Janvier15b}, $\lAsh$ therefore defines where the shock is crossed.  The same is true for $\lAmc$ (\fig{schema3D}) with the MC axis shape determined by \citet{Janvier13}: it provides an estimation of the location where the FR is crossed along its axis. Then, one can derive how physical quantities typically change along the MC axis and the shock shape from sets of MCs and ICME shocks.

In summary, generic dependence of physical parameters across and along the FR can be explored statistically using $\pmc$ and $\lAmc$, while $\lAsh$ allows to explore the dependence away from the shock apex independently of the MC axis direction.  Their results are independent from each other, since MC and shock data have been obtained separately, and the techniques used to derive the FR axis and shock normal are also unrelated. However, we expect that the results are linked to each other from a physical point of view: exploring $\lAsh$ is almost equivalent to exploring both $\pmc$ and $\lAmc$ parameters but without knowing the MC axis direction (this approach is less reliable for the side extensions of the shock where there are no FRs behind).

While $\lAsh$ is positive by definition, both $\pmc$ and $\lAmc$ are signed quantities (indicating on which side the FR is crossed). 
However, as we are limited by the number of MCs, we report the results in function of $|\pmc|$ and $|\lAmc|$, so as to group the FR sides (respectively the FR legs) together and increase the statistics. 

 The statistical study of a parameter in function of $\pmc$, $\lAmc$, or $\lAsh$ can reveal if this parameter has a global dependence across or along the MC axis or along the ICME front within the limit of the statistical dispersion (its standard deviation).  More precisely it can reveal a monotonous variation or one dominated by a positive or negative variation on the scale of the range of $\pmc$, $\lAmc$, or $\lAsh$. However, the correlation coefficients and the slope (of the fitted straight line)
cannot reveal neither significant variations on scales much smaller than the scale of the range studied, nor a more complex relation than the one given by a non-linear function (e.g. a relation with multi-branches). A summary of statistical tests is presented in Appendix~\ref{sect_appendix}.

Finally, it is also worth noticing that the angles $\lAmc$ and $\lAsh$ have large errors, of the order of $20 \degree$, as deduced by comparing the results obtained with the same MCs/shocks but with different methods to deduce $\lAmc$ and $\lAsh$ \citep[See Fig. 2 of][]{Janvier15b}. The impact parameter has also large uncertainties \citep{Lepping10,Demoulin13}.  These large error bars of individual cases are decreased by the statistics, and indeed compatible FR axis and shock shapes were  found by \citet{Janvier15b} from a statistical method while in one third of the same events the difference of relative orientation of the FR axis and shock normal was away of the expected (orthogonal) orientation by more than $25\degree$ \citep{Feng10}.

\begin{figure*}  
\centering
\includegraphics[width=\textwidth,clip]{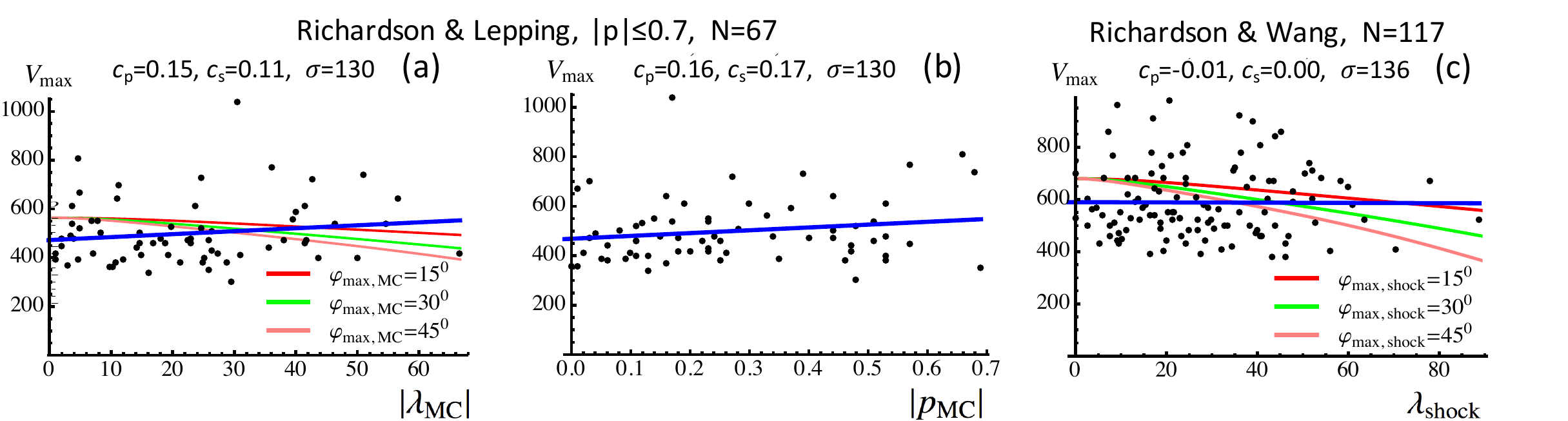}
\caption{Variation of the maximum radial velocity, $\Vmax$, measured after the sheath within the ICME.  The dependence is shown (a) along the MC axis ($|\lAmc|$), (b) across the MC axis ($|\pmc|$), and (c) along the shock ($\lAsh$). \commonCap\  From the graphs, it can be seen that $\Vmax$ is independent of the crossing location.   The red, green and pink lines indicate the expectation of a self similar expansion model (see \sect{Sheath_along}).   
}
\label{fig_lA_p_Vmax}
\end{figure*}

\begin{figure*}  
\centering
\includegraphics[width=\textwidth,clip]{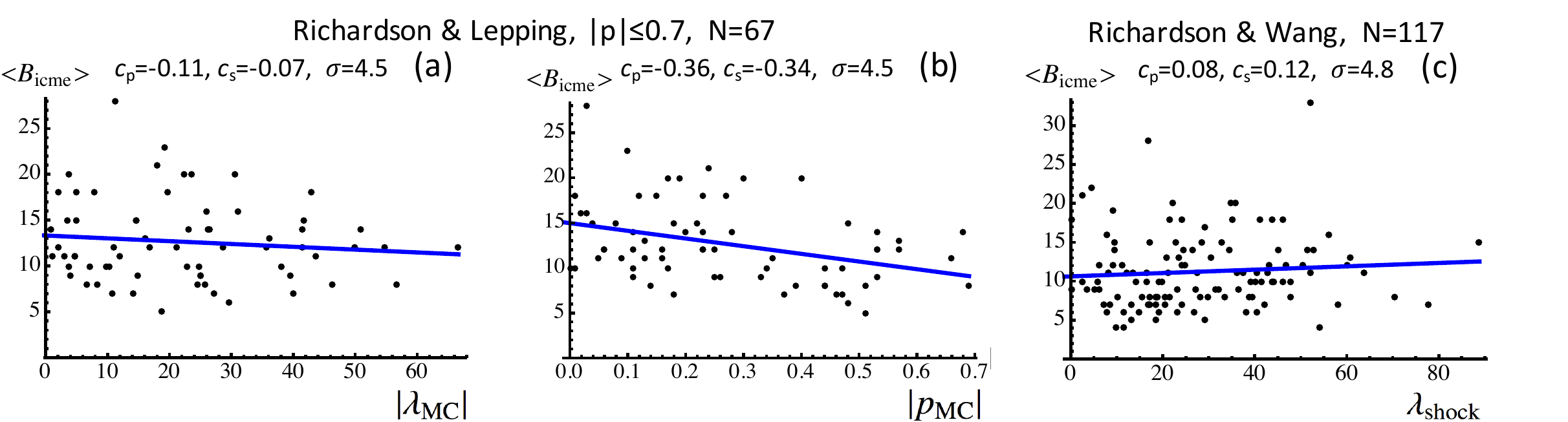}
\caption{Variation of the mean field strength, \Bicme, measured after the sheath within the ICME.  The dependence is shown (a) along the MC axis ($|\lAmc|$), (b) across the MC axis ($|\pmc|$), and (c) along the shock ($\lAsh$). \commonCap\  \Bicme\ is in average uniform  along the MC axis ($|\lAmc|$) and along the shock front ($\lAsh$), while \Bicme\ decreases across the MC axis ($|\pmc|$), as expected (stronger magnetic field in the MC core).
}
\label{fig_lA_p_Bicme}
\end{figure*}

\subsection{Property variations along the ICME front} 
\label{sect_Sheath_along}

%

The sheath thickness, $\Ssh$, shows a large variability for all $|\lAmc|$, $|\pmc|$ and $\lAsh$ values (\fig{lA_p_Ssh}). All the correlation coefficients are low ($\leq 0.11$).
We conclude that the sheath thickness has no global trend along the shock front within the limit of the observed standard deviation ($\approx 0.1$ AU). Some individual ICME sheath may still present variations along the shock front, simply they are not generic for the majority of events.  This result is compatible with the similar mean shape found for MC axis and shock normals \citep{Janvier15b}, see \sect{ICME_generic_model}.

Next, we explore the global properties of ICMEs, such as the maximum, $\Vmax$, and mean velocity, $\Vmean$, computed after the sheath within each ICME.
We found no correlation of $\Vmax$ (\fig{lA_p_Vmax}) and $\Vmean$ (not shown) with any of the location parameters.   Next, we compare this with the expectation of a self-similar expansion model \citep[\eg\ ][ and references therein]{Mostl15} where the shape (axis or shock surface) is simply rescaled with a function of time $f(t)$, so $r(\lA,t)=f(t)\, r(\lA,t_0)$ where $r(\lA,t)$ is the distance to the Sun at $\lA$ and time $t$, and $t_0$ is a reference time. Then the radial velocity is: 
  \BE   \label{eq_self-similar}
  V(\lA,t) = f'(t) \; r(\lA,t_0) \,.
  \EE
We use the elliptical model of \citet{Janvier13} with $b/a=1.3$, a value very close to the ones deduced from the distributions of $\lA$ both for MCs and shocks.  For a given time, $t$, or a velocity weakly dependent of time around the spacecraft location ($f'(t)\approx$ constant), the self-similar expansion implies a decreasing function $V(\lA ,t)$ of $\lA$ for any $\phimax$ value.  This decrease is not observed (\fig{lA_p_Vmax}a,c) and it implies a lack of exact self-similar evolution.

We interpret the above differences as follows. Away from the Sun, the main force acting on the ICME is the drag force which typically depends on the square of the velocity difference between the ICME and the encountered solar wind \citep[\eg , ][ and references therein]{Vrsnak13}.  Then, the drag force tends to make the values of the velocity closer to the one of the encountered plasma, so that the dependence $V(\lA)$ becomes weaker with distances if the MC is embedded in the same type of solar wind.  
Finally, from the results of \fig{lA_p_Vmax}, and the same ones for $\Vmean$, we conclude that the drag force was efficient enough during the travel to 1~AU to uniformize the ICME velocities in the ortho-radial directions. 

Next, we analyse the mean magnetic field strength, \Bicme, computed after the sheath within each ICME.  There is no significant dependence of \Bicme\ with $|\lAmc|$ and $\lAsh$ (\fig{lA_p_Bicme}a,c).  However, \Bicme\ decreases with the impact parameter (\fig{lA_p_Bicme}b) as expected in FR models since the field strength decreases away from the axis due to the inward magnetic tension.  This provides a test for the statistical method: while there is a large variability of the quantities within the ICME cases, the statistical approach still shows a dependence when it is present, a result supported by the tests summarized in Appendix~\ref{sect_appendix}. 

Finally, when the magnetic field strength, $\Baxis$ on the FR axis is considered (derived from the fitted FR model to the \insitu\ magnetic field data), we found only weak correlations with both $|\lAmc|$ and $|\pmc|$ ($\cp$ and $\cs$ are between $-0.2$ and $-0.1$).  The low correlation with $|\lAmc|$ is in agreement with the result of \Bicme\ (\fig{lA_p_Bicme}), but limited to the MC part of the ICMEs.  The low correlation of $|\pmc|$ with $\Baxis$ is expected if there is no bias in the results of the fitted FR model.
   
In conclusion, with a statistical approach on large sets of MCs, shocks and ICMEs and with the estimated parameters $|\lAmc|$, $|\pmc|$ and $\lAsh$ we are able to explore the behavior of global ICME quantities in directions orthogonal to the spacecraft trajectory.   This new approach provides 
new results such as a sheath thickness and an ICME radial velocity which show no global and systematic dependence along the ICME front within the limit of the observed standard deviation, \ie\ there is no dependence which would be present in the majority of ICMEs. A priori we cannot exclude, \eg , a situation where there are several groups of ICMEs with different dependencies which cancel on average.  However, we explored this by dividing the ICME set in few groups with common characteristics (\eg\ velocity, field strength or size) but we did not find any sub-group with specific properties.

\section{Quantitative generic ICME structure} 
\label{sect_ICME}

\subsection{Relationship between the axis and shock angular extensions} 
\label{sect_ICME_phimax_A_S}
   
In previous studies \citep{Janvier13,Janvier15b}, two parameters could not be determined from \insitu\ data: the maximum angular extension of the structures, $\phimaxmc$ and $\phimaxsh$.  
However, under the assumptions that all ICMEs contain a FR and that all shocks are driven by ICMEs, we can use the relative number of detected FRs to that of shocks in a given time interval, to constrain $\phimaxmc$ in function of $\phimaxsh$, as follows.  For a FR with a constant radius $R$ along its axis, its cross section, as seen from a spacecraft located at a distance $D$, is
approximatively the product of its diameter and its projected length $2~\phimaxmc ~D$ (we suppose that $R<<D$ so that the cross section is a thin elongated band of the sphere of radius $D$).  
For a FR with a variable radius along its axis, its mean radius $\Rmean$ replaces $R$. Then the FR cross section seen by a spacecraft is
  \BE \label{eq_CS_FR}
  CS_{\rm FR} \approx 4~\phimaxmc ~\Rmean ~D \, .
  \EE

The cross section of a shock with an ellipsoid shape, \eq{f}, is:
  \BE \label{eq_CS_S}
  CS_{\rm shock} \approx \pi ~b~c \approx \pi  ~b/c ~D^2 ~\sin ^2{\phimaxsh} \, .
  \EE
We choose to define above $\phimaxsh$ with the largest extension $c$ since we estimate $\phimaxsh$ below from the imager data.  The 2D images do not provide the orientation of the FR, nor from which point of view the sheath is observed so there is no indication whether the extension seen is closer to $b$ or $c$.
With a uniform distribution of the elongation orientation the measured width is closer to the longest extension.   What is estimated from the imagers is $c=b$ for the axisymmetric case and $2 c/ \pi$ for the limit case of $c>>b$.   Then, we define $\phimaxsh$ with $c$ in \eq{CS_S} as the best approximation since $c/b$ is unknown.  Furthermore there is the caveat that the shock extension is rarely traced in CME images, so the observed angular extension is rather the extension of the dense sheath.

Because of the broad range of latitudes of the solar sources and the solar rotation, CMEs are randomly launched in a nearly uniform range of directions for a spacecraft observing at a fixed position for years. Then, the expected ratio of the number of FRs, ${\rm N}_{\rm FR~with~shock}$, to the number of ICME front shocks, ${\rm N}_{\rm shock}$, is the ratio of their cross-sections. This provides a relationship between $\phimaxmc$ and $\phimaxsh$:
  \BE \label{eq_phi_max_axis}
  \phimaxmc \approx \frac{\pi}{4} \frac{{\rm N}_{\rm FR~with~shock}}{{\rm N}_{\rm shock}}
  \frac{D}{\Rmean} \frac{b}{c} ~\sin ^2{\phimaxsh} \, .
  \EE

\begin{figure}  
\centering
\includegraphics[width=0.4\textwidth,clip]{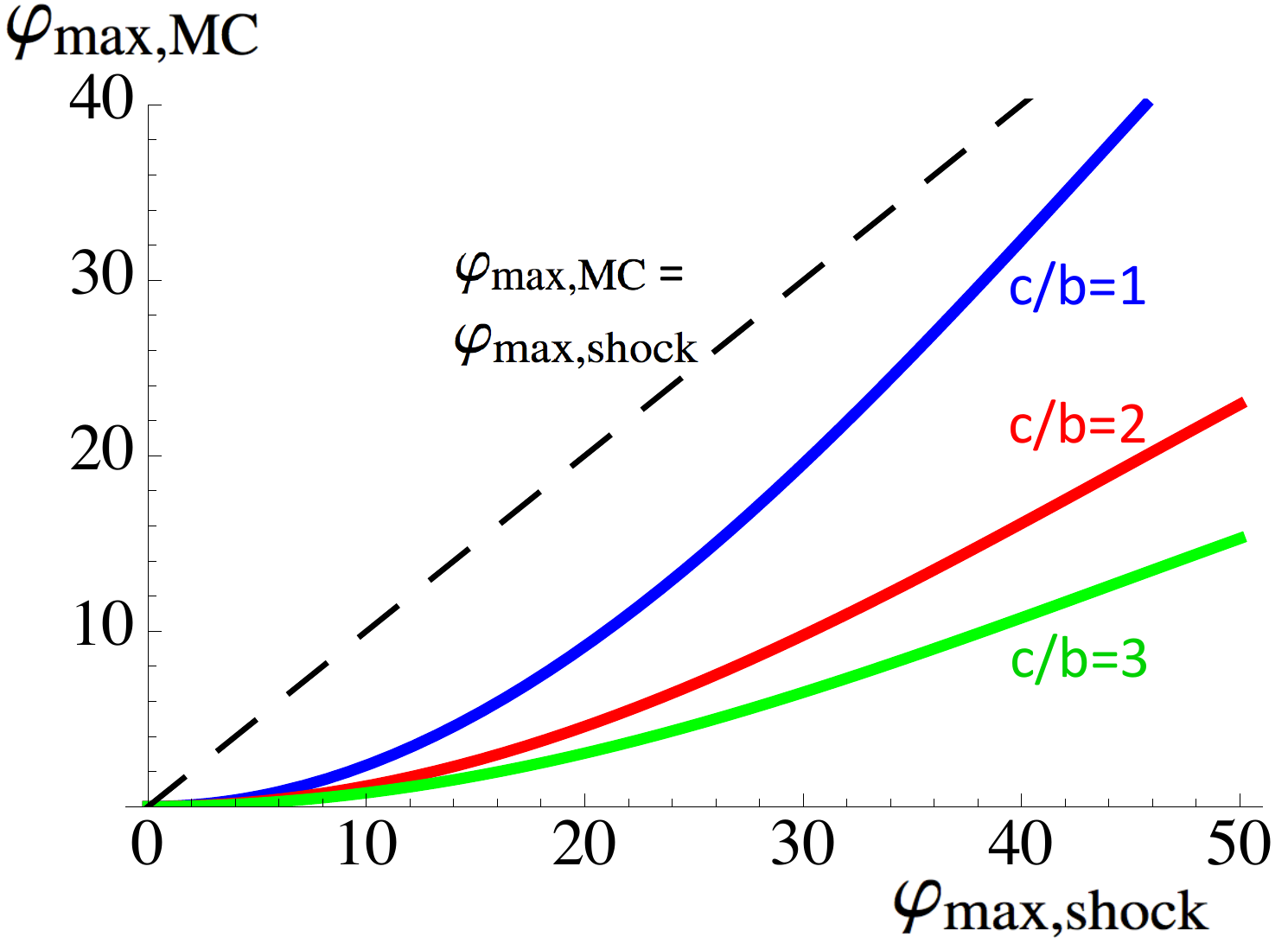}
\caption{$\phimaxmc$ in function of $\phimaxsh$ from \eq{phi_max_axis}. 
${\rm N}_{\rm FR~with~shock}$ is estimated with the number of detected MCs (45) and ${\rm N}_{\rm shock}$ is the total number of front ICME shocks (216) detected during the same period of time.  The curves are for the shock-surface aspect ratio $c/b$ defined in \fig{schemaCart}.
}
\label{fig_phimaxA_S}
\end{figure}

\citet{Janvier14b} associated 117 shocks to ICME sheath fronts from Richardson \& Cane ICME list and Wang et al. shock list (see \sect{Sheath_association}). Then, 99 other shocks were also considered. These shocks were not directly associated with ICME sheath fronts, but were also not found within ICME sheaths, within an ICME or within 6 hours after an ICME. The authors argued that most of these remaining shocks were actually associated with ICMEs not detected \insitu. Two reasons can be given. First, the frequency of shocks driven by stream interaction regions (SIRs) is much lower than the frequency of shocks driven by ICMEs at 1~AU, and second, the distribution of $\pobsl$ for those 99 shocks is similar to $\pobsl$ for shocks with a detected ICME behind.  This result is also in agreement with the expectation that the shock lateral extension is broader than the driving ICME.  In summary, we estimated a total of 216 shocks  detected in front of ICME sheaths in the time interval 1998-2008.  
  
  Within the same time interval, 45 MCs and 36 MC-like events were detected.   Then, the number of FRs entering in \eq{phi_max_axis} is between 45 and 81.  Next, the mean FR radius for MCs is $\Rmean \approx 0.12$~AU \citep{Lepping10}.
Although $\phimaxsh$ cannot be determined by the shape of $\pobsl$, we constrain $\phimaxsh <60 \degree$ following the analysis in Section 3.4 of \citet{Janvier15b}.
Moreover, the mean half angular width of a limb CME is $\approx 30 \degree$ \citep{Wang11}.  This angular width, corresponding to the CME sheath, is typically kept as CMEs propagate away from the Sun, so $\phimaxsh$ is around $30 \degree$.

With the above set of parameters, $\phimaxmc$ is estimated for a given $\phimaxsh$. 
In \fig{phimaxA_S} we show the conservative case where only MCs are considered for the number of FRs.  For the axisymmetric case, $b=c$, the shock is typically broader by about $10 \degree$ than the MC axis.  As $c/b$ increases the shock becomes even broader by $20 \degree$ to $30 \degree$. From the impact parameter distribution, \citet{Demoulin13} showed that the typical FR cross section is indeed not circular but flatten in the radial direction by a factor ranging from $1.5$ to $3$ depending on the FR model selected. Taking into account this flattening would increase $\Rmean$ value used above by such factor.  It has the same effect as increasing $c/b$ as discussed above. Next, if we consider MCs and MC-like events, the estimation of $\phimaxmc$ needs to be multiplied by $81/45 =1.8$, so $\phimaxmc$ and $\phimaxsh$ are much closer, with the possibility of a shock extending less than the MC axis ($\phimaxsh -\phimaxmc$ is in the range $[-10 \degree, 20\degree]$).   Finally, if we consider that only 70\% of the shocks observed at 1~AU are driven by ICMEs \citep{Jian06b}, the estimation of $\phimaxmc$ is multiplied by $1/0.7 \approx 1.43$ so it has the same effect, but weaker, than including the MC-like events.
 
\begin{figure}  
\centering
\includegraphics[width=0.5\textwidth,clip]{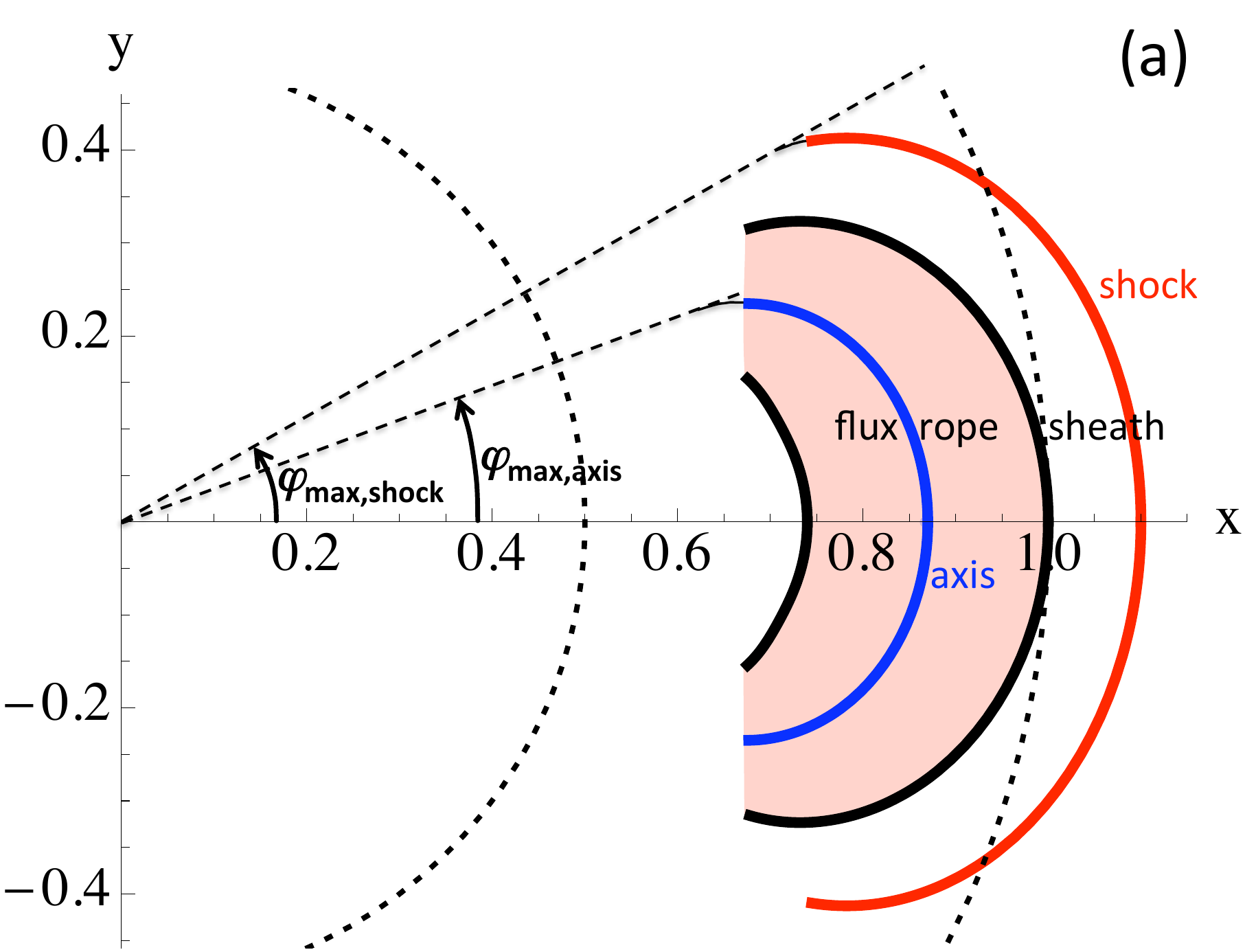}
\includegraphics[width=0.5\textwidth,clip]{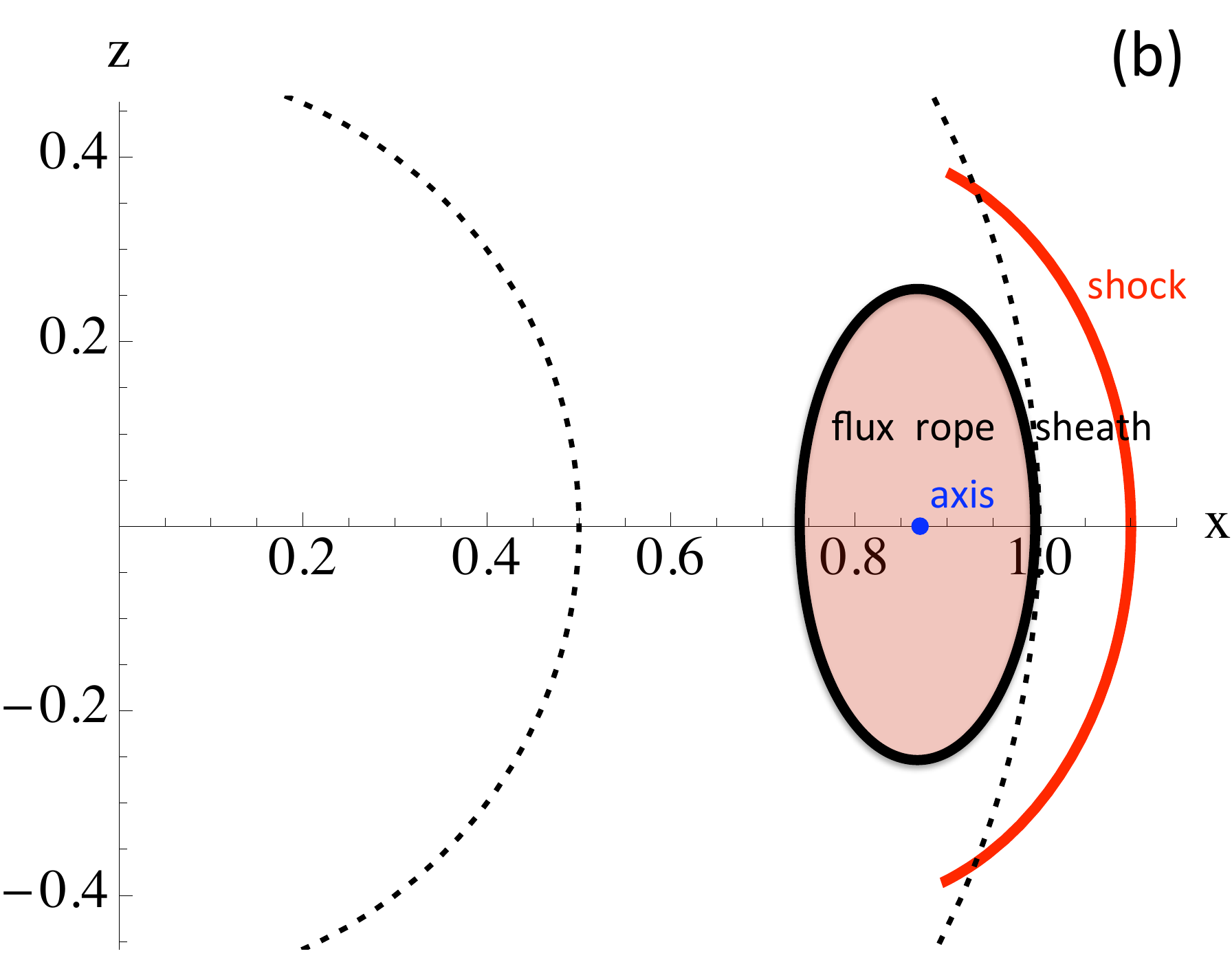}
\caption{Summary of the results of the axis and shock shapes providing a quantitative version of the sketch drawn in \fig{schema3D}. {{\bf (a)} Cut in the plane of FR axis and {\bf (b)} cut in the plane orthogonal to axis plane at the apex.} $\phimaxsh$ is the {only free} parameter, set here to $30 \degree$.  $\phimaxmc = 20 \degree$ is estimated from \eq{phi_max_axis} with $N_{\rm FR}=45$ (\fig{phimaxA_S}). The aspect ratio $\bSa =1.2$ and $1.3$ for axis and shock, respectively, are mean values of the results of \citet{Janvier14b}. 
At the apex, the sheath thickness is set to the mean value of 0.1~AU, {however it is very case dependent} (\fig{lA_p_Ssh}). The radial FR extension is deduced from \citet{Janvier14b}. The FR cross section is elliptic with an aspect ratio of 2 as estimated by \citet{Demoulin13}.
The FR front is set at 1~AU, with the axis and shock shapes rescaled by their apex distances ($0.87$ and $1.1$~AU respectively). 
The two dotted lines mark the radial distance to the Sun at 0.5 and 1~AU.}
\label{fig_final}
\end{figure}

\subsection{Quantitative generic ICME structure} 
\label{sect_ICME_generic_model}

With the quantitative information found above and in previous studies on the different parts of ICMEs (shock, sheath, FR axis), we can derive a quantitative generic model of ICMEs based on the analysed \insitu\ observations, as follows. Since both the imager and \insitu\ data are not sufficient to define the aspect ratio of the shock surface in ortho-radial directions, we keep below the simplest solution compatible with the observations: an axisymmetric configuration.   The global ICME configuration can be derived with other aspect ratios, as well as other $\phimaxsh$ value, with the same procedure.  

   The generic shapes of the MC axis and shock shells have been determined with the ellipsoidal model, which provides the closest distribution $\Pl$ from the observed one, $\pobsl$, both for MCs and shocks \citep{Janvier15b}.   While errors in the determination of MC axis and shock normals are large for individual events, we have shown that these errors have a small effect on the determination of the generic shapes. 
We set $\phimaxsh=30\degree$ and, with a conservative approach, we include only MCs as detected flux ropes.  This implies $\phimaxmc \approx 20\degree$ (\fig{phimaxA_S}). 
   
First, we consider the plane containing the FR axis (\fig{final}a).  The FR radius of MCs, from Lepping's list, was found to depend on $\lA$ \citep{Janvier14c}. We use the approximation $R(\lA)=0.13 - 10^{-5} \lA^2$~AU for this dependence. This radius is added/substracted in the direction orthogonal to the axis to derive the front/rear of the FR, respectively (\fig{final}a).  This means that in the present model, we do not take into account the amount of reconnected flux, \ie\ the back region, that is observed in some flux ropes.  We normalize the front apex to 1~AU.  
   The sheath thickness, $\approx 0.1$~AU (\sect{Sheath_global}, \fig{lA_p_Ssh}), is added radially to the FR front apex to define the shock apex.  Then, the shock shape found in \citet{Janvier14b}, rescaled to the sheath apex, is added.  Finally, we note that all observations were made at 1~AU while \fig{final} shows the FR, sheath and shock around 1~AU. To draw \fig{final} we suppose that their shapes are not significantly modified around 1~AU (only rescaled by a factor so that $\Pl$ is not changed).  This hypothesis is stronger for large $\lA$ values where the number of detected cases are small so the axis and shock shapes are less constrained by observations for larger $\lA$ values.  

We show an elliptical shape for the FR cross section with an intermediate aspect ratio of $2$ in \fig{final}b.  
We included the sheath thickness at the apex as explained above, as well as the same shock shape (supposing axisymmetry).
  
A configuration qualitatively comparable to \fig{final}a was first derived from multi-spacecraft crossings of a single MC by \citet{Burlaga90}. But such favorable observations with multi-spacecraft crossings remain rare as the number of spacecraft is at most a few. Moreover, the various methods to reconstruct the FR from the \insitu\ data are prone to large uncertainties.  Indeed, recent studies led by different authors have reached different conclusions on the axis shape even when analysing the same multi-spacecraft data of a MC observed with favorable conditions.  For example, the opposite results of \citet{Farrugia11} and \citet{Ruffenach12} on the axis shape are compared in Fig.~3 of \citet{Demoulin14}. 
In contrast, our statistical approach allows to minimize the present uncertainties on the axis and shock normal directions and to derive a quantitative generic configuration for ICMEs (\fig{final}).

The result of \fig{final} is also qualitatively comparable to the expected features of coronal mass ejections as frequently sketched [\eg\ \citet{Zurbuchen06}, and references therein]. However, our model, directly derived from observational data, differs quantitatively in the shapes of the different ICME components.  

We emphasize that the FR axis and shock shapes are consistent with each other in \fig{final} while they are derived from different \insitu\ data and modelling.  Furthermore, the typical FR cross-section shape, the FR size, the FR axis and shock shapes combined together imply a sheath thickness which is nearly independent of $\lAsh$ just as found from a direct analysis of observations (\fig{lA_p_Ssh}).  This is compatible with the assumption of axisymmetry around Sun-apex line used in our analysis of the shock shape.  In other words, around the apex, the two principal curvatures of FR boundary are comparable,
then the FR obstacle drive in front a shock with an approximative axi-symmetry around the FR mean motion direction.

\section{Conclusion} 
\label{sect_Conclusion}

Imager data only provide a 2D distribution of mass density integrated along the line of sight. In situ measurements typically provide the magnetic field and plasma parameters along the 1D trajectory of the crossing spacecraft while the crossing of the same ICME by several spacecraft is rare.  Then, our understanding of the 3D structure of ICMEs is partial.  

In this study, we further develop a statistical approach of the \insitu\ data in order to derive information on the 3D shape of ICMEs and their components.  Since it uses single crossings of a large number of ICMEs, the results can only provide a generic description of the 3D shape, washing out all the peculiarities of single events.  

We first investigate the generic shape of the front shock.  In previous studies \citep{Janvier14b,Janvier15b} we derived its generic shape assuming a symmetry around the line from Sun to shock apex, and we showed that an ellipsoid shape was the best fit model, of the ones tested, to the observed distribution $\mathcal{P}(\lAsh)$ of the location angle $\lAsh$ (defined in \fig{schema3D}). 
Here, we tested how an asymmetric shock shape is compatible with the observed distribution of $\lAsh$ by computing the distributions generated by a shock with a 3D ellipsoid shape.  We show that a larger elongation of the ellipsoid in one direction can be partly compensated by a more bent shape in the orthogonal direction for the computed $\lAsh$ values, then the observed $\mathcal{P}(\lAsh)$ is compatible with shock shapes with an aspect ratio (in the two orthogonal directions across the propagation direction) lower or equal to 3, but not significantly larger. 
Then, the \insitu\ data constrained the asymmetry around the Sun-apex line without determining it.

In numerical simulations where the ICME apex is directed into a dense and slow solar wind while its sides are in a fast solar wind, the shock shape is bended around the apex up to reversing the curvature: the shock surface has a dip around the apex.  In this study we test the implication of such dipped shock surface on $\mathcal{P}(\lAsh)$.
A flattening of the shock front has already a large effect on $\mathcal{P}(\lAsh)$ since low $\lAsh$ values are more frequent as the shock front is flatter.  Further, as the dip of the surface is stronger, this effect is present at larger $\lAsh$ values. Then, flat and even more dipped shock surfaces cannot be generically present in ICMEs otherwise $\mathcal{P}(\lAsh)$ would be different than observed.         

Next, the location angle $\lAsh$ provides an estimation of where the shock was crossed by the spacecraft ($\lAsh=0$ at the apex and $\lAsh$ is a growing function of the distance to the apex).  
In the same way $\lAmc$ (see \fig{schema3D}) provides an estimation of where the spacecraft is crossing the MC along its axis, while the impact parameter $\pmc$ provides a comparable information away from the axis.  
Then, these quantities provide two independent ways, with shock and MC data, to localize the spacecraft trajectory within the encountered ICME.  However, because of the large uncertainties on these three parameters this information can only be used on a statistical basis.  
This approach allows the derivation of the generic variations of global ICME quantities both along the shock front and the MC axis, as well as across the axis.

We found that the sheath thickness: 1) has a mean value of 0.1~AU (comparable to other studies), 2) has a large variability between cases and 3) has no global dependence along the FR/shock front within the limits of the observed standard deviation (\fig{lA_p_Ssh}). 
Third, the maximum and mean velocities (along the spacecraft crossing) are also almost constant along the shock front contrary to the expected results of a self-similar expansion model.
These uniform velocities are interpreted as the consequence of a strong coupling to the ambient solar wind by the drag force.   Finally, we found that the average ICME magnetic field strength is uniform along the FR but decreases across it as expected by FR models.

With all the above results, we finally considered the generic shape of the whole ICME.  The ellipsoidal model has the best fit to the observed distributions $\pobsl$. It defines 
the axis and shock generic shapes up to {the free angular extensions $\phimaxmc$ and $\phimaxsh$ (defined in \fig{schema3D})} as $\Pl$ of the ellipsoidal model has only a small dependence on this parameter \citep{Janvier14b}.  We relate $\phimaxmc$ to $\phimaxsh$ using the observed numbers of ICME shocks with and without associated MCs.  Next, $\phimaxsh$ is {approximated} by the mean angular extension of CMEs observed close to the solar limb (so minimizing projection effects, {but with the caveat that the sheath, rather than the shock, is observed}).  Finally, the mean variation of the flux tube radius with $\lA$ and the typical shape of the FR cross section are taken from previous studies \citep[][ respectively]{Janvier13, Demoulin13}.  
By combining all these results, we established a quantified generic ICME structure comprising a flux rope, a sheath and an associated shock (\sect{ICME}).  All the parts of this generic model, shown in \fig{final}, are constrained by observations.   The results are in the line of some previous studies based on multi-spacecraft observations of one event \citep[\eg][]{Burlaga90,Ruffenach12}, with the advantage that using a large set of MCs/shocks crossed at various $\lA$ allows to derive a quantitative and generic model. As far as we know, this is the first time that a quantitative and 
generic 3D-shape of ICMEs and their sub-structures is derived from a 
statistical analysis.  The derived ICME model can have several applications such as the examples described below.

A first application of our results is to provide constraints for numerical simulations of ICMEs.  In particular the quantitative generic shock, sheath and axis derived (\fig{final}) provide a landmark to compare with numerical simulations.
Furthermore, the shock and FR axis distributions of $\Pl$ of these simulations can be compared with the observed ones \citep[as derived in Figure 4 of][]{Janvier15b}. 
   
Another possible application of the results presented here is linked with the transport of energetic particles in the interplanetary medium.
The most energetic solar particles produced during a flare can travel in interplanetary structures near or inside ICMEs \citep[\eg][ and references therein]{Masson12}. 
Then, to improve the knowledge of the transport properties of these solar energetic particles (\eg\ their time of travel) it is necessary to know the global shape of the ICME magnetic structure, as well as the length of the FR field lines \citep{Demoulin16}.

More global implications are related with depletions of galactic cosmic ray fluxes arriving at Earth in association with the passage of an ICME \citep[\eg ][]{Masias-Meza16}.  Two physical mechanisms are typically thought to explain the depletion of cosmic ray fluxes (which occur in two steps): the deflection of particles by the more intense ICME magnetic field and by the turbulence properties at the ICME shock (the so-called diffusive barrier, \eg\ \citealp{Cane00}).
The quantification of these two effects require the knowledge of the global structure of ICMEs. 

Finally, another application of our results is  for space weather. 
The results summarized in \fig{final} could be used to develop the ellipse evolution model of \citet{Mostl15} by incorporating a typical FR and sheath.   Depending on the available observations the parameters of the model could be specific to the event studied or the derived generic values found above can be used.

\begin{acknowledgements}
 We thank the referee for her/his comments which improved the manuscript.
M.J. acknowledges fundings from the Northern Research Partnership with travel support to the Observatoire de Paris.
S.D. acknowledges partial support from the Argentinian grants UBACyT 20020120100220 (UBA), PICT-2013-1462 (FONCyT-ANPCyT), PIP-CONICET-11220130100439CO, 
and PIDDEF 2014-2017 number 8.
This work was partially supported by a one-month invitation of P.D. to the Instituto de Astronom\'ia y F\'isica del Espacio, 
and by a one-month invitation of S.D. to the Observatoire de Paris.
S.D. is member of the Carrera del Investigador Cien\-t\'\i fi\-co, CONICET.\end{acknowledgements}
 
\bibliographystyle{aa}
\bibliography{icme_3D}
\IfFileExists{\jobname.bbl}{}
{\typeout{}
\typeout{****************************************************}
\typeout{****************************************************}
\typeout{** Please run "bibtex \jobname" to optain}
\typeout{** the bibliography and then re-run LaTeX}
\typeout{** twice to fix the references!}
\typeout{****************************************************}
\typeout{****************************************************}
\typeout{}
}

\Online
\begin{appendix} 

\section{Statistical tests}
\label{sect_appendix}

\begin{figure*}  
\centering
\includegraphics[width=0.8\textwidth,clip]{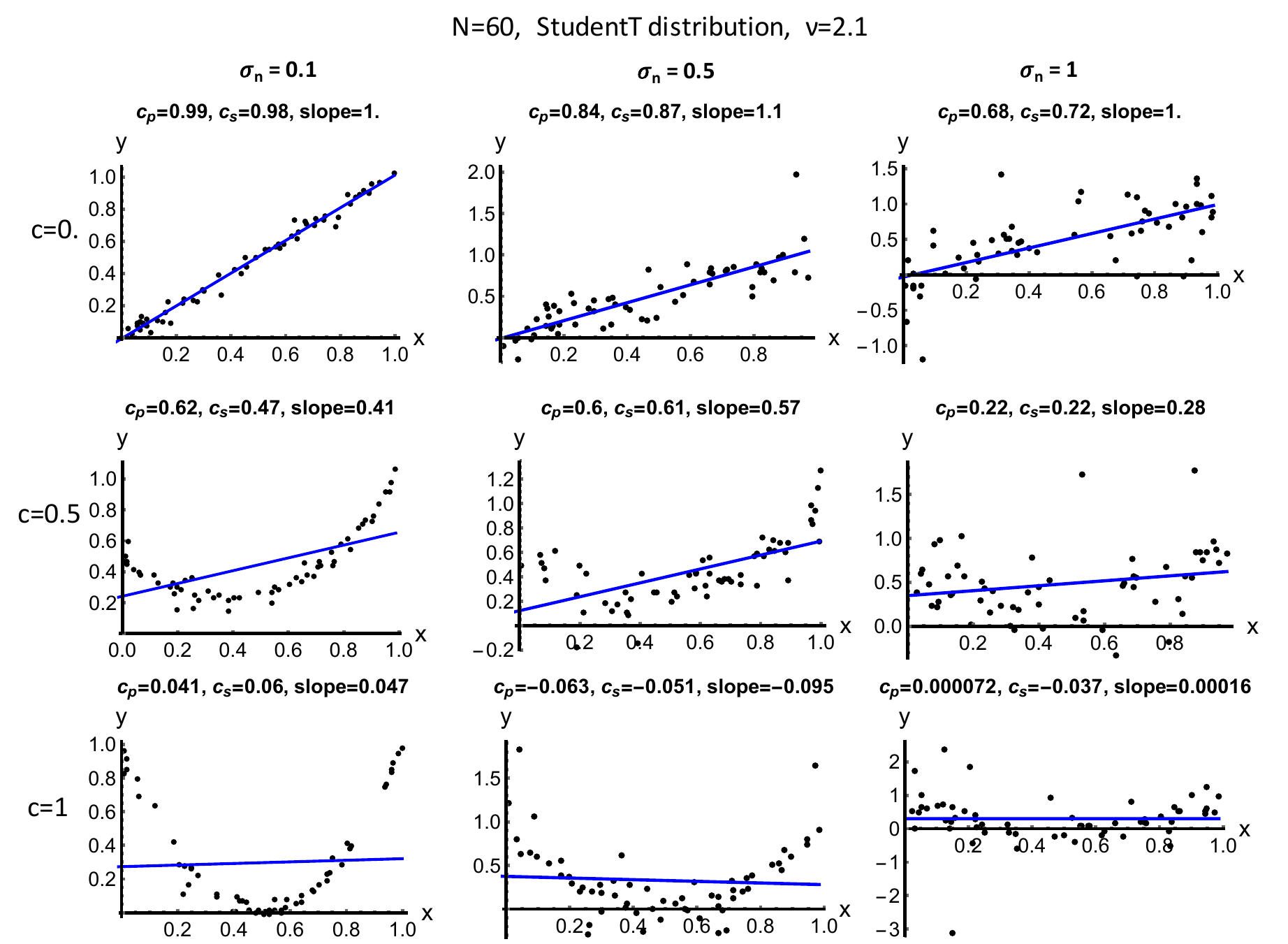}
\caption{Examples for the 3 cases $c=0,0.5,1$ (from top to bottom) of \eq{y(x)} with the added noise with the studentT distribution ($\nu =2.1$) and for 3 levels of noise (standard deviation $\sigma_n= 0.1, 0.5, 1$ from left to right column). In each panel $N=60$ points are selected with a different random seed. The blue straight line is the least square fit to the simulated points (in black). The correlation coefficients and the slope of fitted straight line are written on top of the panels.   
}
\label{fig_model_fit}
\end{figure*}

\begin{figure*}  
\centering
\includegraphics[width=0.8\textwidth,clip]{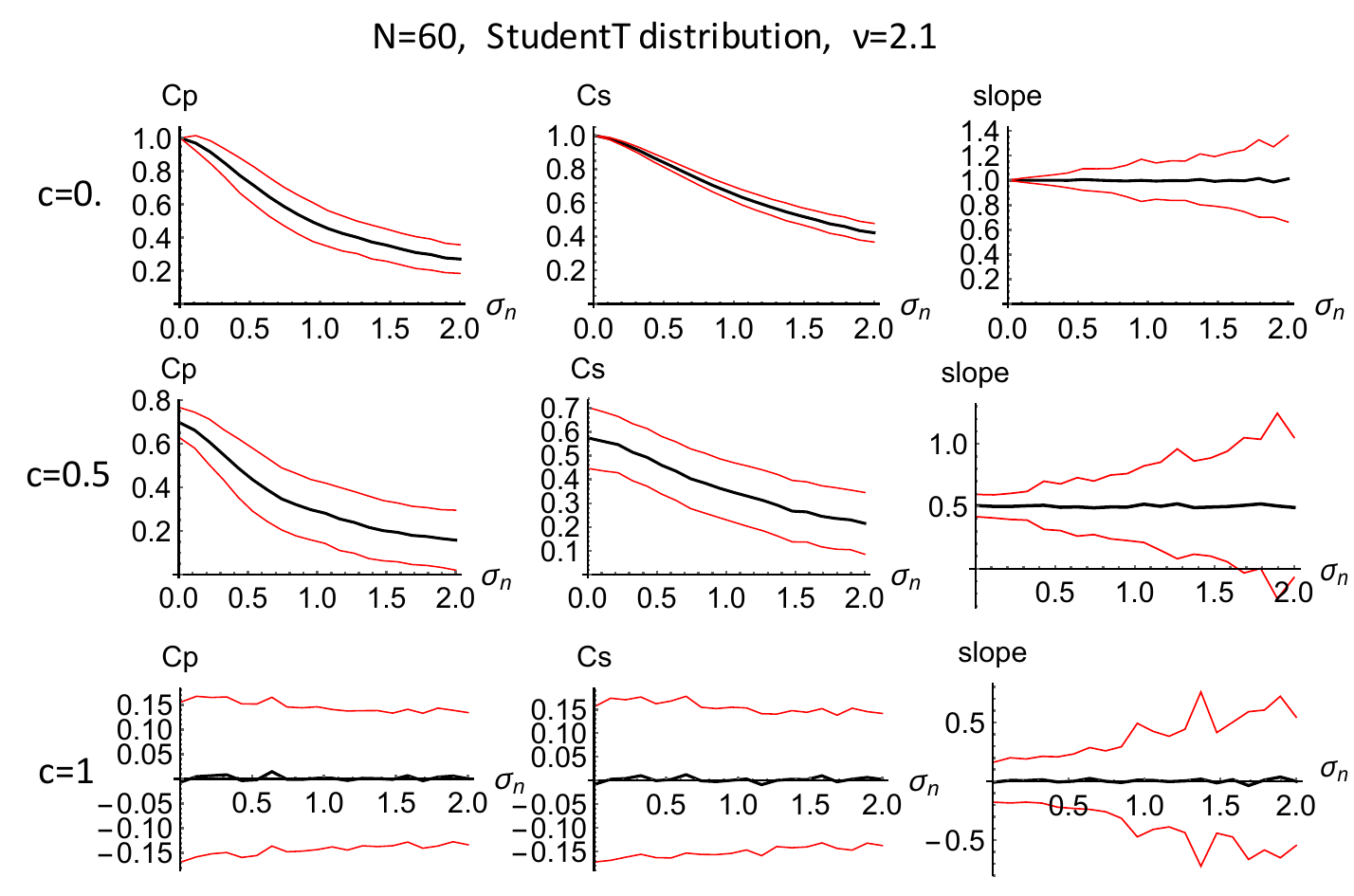}
\caption{Statistical results for the 3 cases $c=0,0.5,1$ (from top to bottom) of \eq{y(x)} and the added noise with the studentT distribution ($\nu =2.1$).
The correlation coefficients and the slope of fitted straight line are plotted in function of the standard deviation $\sigma_n$ of the added noise.
The tests, shown in \fig{model_fit}, are repeated 1000 times with a modified random seed.  The means are shown with black lines and the error range, defined with the standard deviation, is shown with the red lines. 
}
\label{fig_model_stat}
\end{figure*}

   We analyzed the correlation coefficients and the linear least square fits of various parameters in \sects{Sheath_method}{Sheath_along} with a limited number of cases (between 59 and 117) and with significant dispersions.   In order to better characterize the results we test below the same statistical tools on models with added noise.  
   
With the limited number of available data and their dispersion we cannot hope to derive more than global tendencies.   Then, we consider a polynomial function $y_{\rm mod}(x)$ to test the robustness of the Pearson and Spearman correlation coefficients, $\cp ,\cs$, and of the slope of the linear fit.   Since these three coefficients are insensitive to the mean of $y_{\rm mod}(x)$ we skip the constant of the polynomial.  We define the range of $x$ variations to the normalized interval $[0,1]$.  The $y$ variable is also normalized so that $y(1)=1$.  We limit ourselves below to a polynomial of second order as it is sufficient to show the strengths and limits of the analysis.    These constraints define the model function  
  \begin{equation}
  y_{\rm mod}(x) = (1 - c) \, x + c \, (2 x - 1)^2
  \label{eq_y(x)}
  \end{equation}
where $c$ is a free coefficient.    For $c=0$, $y_{\rm mod}(x)$ is linear while for $c=1$, $y_{\rm mod}(x)$ is a parabola symmetric around the straight line $x=1/2$.   

Next, we defined $N$ values of $x$ randomly selected in the range $[0,1]$ with a uniform distribution.  Then, we add noise to $y_{\rm mod}(x)$ with a mean $\mu_{n}=0$ and a given standard deviation $\sigma_{n}$.  We tested three distributions for the random noise: uniform, normal (or Gaussian) and StudentT. They are functions of $\mu_{n}$ and $\sigma_{n}$.  Only the StudentT distribution has a third parameter called $\nu $ and its standard deviation is $\propto \sqrt{\nu/(\nu-2)}$ so it is defined only for $\nu >2$.   The lower $\nu$ is, the more extended the distribution tails are (they behave as $|x-\mu_{n}|^{-\nu -1}$).

In order to illustrate the effect of outsider points, we select cases with the StudentT distribution with $\nu =2.1$ as it has large tails.  Different outsiders are present in the panels of \fig{model_fit} since the random seeds are different.  Comparing with the results of other distributions (\eg\ normal and uniform) with weak or without tails, comparable results are found.  In fact for a given noise magnitude, measured by $\sigma_{n}$, the correlation coefficients decrease less rapidly with $\sigma_{n}$ for lower $\nu $, so for distributions with more numerous outsiders and farther away from the mean, as the core of the distribution needs to be narrower to keep the same $\sigma_{n}$ value. 

The results derived from the examples shown in \fig{model_fit} were checked by running different random distributions and comparing the plots.  Next a statistical analysis was performed by performing $n_{\rm test}$ cases.  The mean and its associated standard deviation of the results are represented in \fig{model_stat}
with $n_{\rm test}=1000$ and again for the StudentT distribution with $\nu = 2.1$, $N=60$, and $c=0,0.5,1$.

  For a linear model, $c=0$, the correlation coefficients detect the linear relation up to $\sigma_n$ values about two times larger than the global variation ($=1$ here).  The slope of the fit is more robust than $\cp ,\cs$ since its mean is close to $1$. Simply its error bar increases linearly with the noise level.
  
   The case with $c=0.5$ is also well detected by the correlation coefficients and the slope, also up to $\sigma_n$ values about two times larger than the global variation ($=0.5$ here).   The cases with uniform and normal distributions show similar curves but with a faster decrease with increasing $\sigma_n$ so that the threshold of $\sigma_n$ is rather comparable to the global variation.
   
 For both cases $c=0$ and $=0.5$ the Spearman coefficient decreases less rapidly than the Pearson one for increasing $\sigma_{n}$ as the ranks, and not the data values, are considered.  The slope is also slightly less dispersed for lower $\nu $.  In contrast, \figss{lA_p_Ssh}{lA_p_Bicme} have much less prominent outsiders, so their effects is even lower.  The outsiders are not an issue in these data.
 
  The case with $c=1$ is an extreme case where no global dependence is present (the positive correlation part cancels exactly the negative correlation part). 
In such a case the correlation coefficients and the slope are unable to detect a dependence.   However, scatter plots such as in \fig{model_fit} do show this dependence up to a noise level $\sigma_n$ comparable to the function maximal excursion.   In such a case, the least square fit by a polynomial of second order (or higher if needed) would show the underlying dependence.  However, since no such dependence is seen in the scatter plots of  \figss{lA_p_Ssh}{lA_p_Bicme} such approach is not pursued.  Still, these data show that there is no global linear trend, up to the limit of the data standard deviation. 
 
   In order to decrease the above limitations more numerous data would be needed. For example repeating the above test with $N=240$,  the results of \fig{model_stat} are reproduced with less fluctuating curves and the error bars (distance between red curves) are about twice lower. Still the mean tendency, shown with the black lines, are almost unchanged.   We conclude that the results in \figss{lA_p_Ssh}{lA_p_Bicme} with $N\approx 60$ and those with $N\approx 120$ have close statistical significance (only a factor $\approx \sqrt{2}$ of differrence).
   
\end{appendix}

\end{document}